# A Public Key Block Cipher Based on Multivariate Quadratic Quasigroups*

Danilo Gligoroski[1] and Smile Markovski[2] and Svein Johan Knapskog[3]

[1] Department of Telematics, Faculty of Information Technology, Mathematics and Electrical Engineering, The Norwegian University of Science and Technology (NTNU), O.S.Bragstads plass 2E, N-7491 Trondheim, NORWAY, danilog@item.ntnu.no

[2] "Ss Cyril and Methodius" University, Faculty of Natural Sciences and Mathematics, Institute of Informatics, P.O.Box 162, 1000 Skopje, MACEDONIA, smile@ii.edu.mk

[3] Centre for Quantifiable Quality of Service in Communication Systems, Norwegian University of Science and Technology, O.S.Bragstads plass 2E, N-7491 Trondheim, NORWAY, Svein.J.Knapskog@q2s.ntnu.no

***Abstract*** – We have designed a new class of public key algorithms based on quasigroup string transformations using a specific class of quasigroups called *multivariate quadratic quasigroups (MQQ)*. Our public key algorithm is a bijective mapping, it does not perform message expansions and can be used both for encryption and signatures. The public key consist of $n$ quadratic polynomials with $n$ variables where $n = 140, 160, \ldots$. A particular characteristic of our public key algorithm is that it is very fast and highly parallelizable. More concretely, it has the speed of a typical modern symmetric block cipher – the reason for the phrase *"A Public Key Block Cipher"* in the title of this paper. Namely the reference C code for the 160–bit variant of the algorithm performs decryption in less than 11,000 cycles (on Intel Core 2 Duo – using only one processor core), and around 6,000 cycles using two CPU cores and OpenMP 2.0 library. However, implemented in Xilinx Virtex-5 FPGA that is running on 249.4 MHz it achieves decryption throughput of 399 Mbps, and implemented on four Xilinx Virtex-5 chips that are running on 276.7 MHz it achieves encryption throughput of 44.27 Gbps. Compared to fastest RSA implementations on similar FPGA platforms, MQQ algorithm is more than 10,000 times faster.

## 1 Introduction

The public key paradigm initially described in the seminal paper of Diffie and Hellman [9], have completely changed and re-shaped modern cryptography. The fruits of that change are noticeable now, 30 years after, with the boom of the Internet, digital telecommunications, the convergence of information and communication technologies and the onset of the modern e-society. Most of the security protocols in these fields in one way or the other use the Public Key paradigm.

The most popular Public Key Cryptosystem (PKC) schemes are the Diffie and Hellman (DH) key exchange scheme based on the hardness of discrete logarithm problem [9], the Rivest, Shamir and Adleman (RSA) scheme based on the difficulty of integer factorization [40], and the Koblitz and Miller (ECC – Elliptic Curve Cryptography) scheme based on the discrete logarithm problem in an additive group of points defined by elliptic curves over finite fields [26,31]. There are two common characteristics of these well known PKCs (DH, RSA and ECC): 1. their speed – which frequently is a thousand times lower than the symmetric cryptographic schemes, 2. their security – which relies on one of two hard mathematical problems: efficient computation of discrete logarithms and factorization of integers.

Several other ideas have been proposed during the last 30 years, such as McEliece PKC based on error correcting codes [30], Rabin's digital signature method [39], PKCs based on lattice reduction problems

* This is an extended and updated version of a paper "Multivariate Quadratic Trapdoor Functions Based on Multivariate Quadratic Quasigroups", Proceedings of the AMERICAN CONFERENCE ON APPLIED MATHEMATICS (MATH '08), Cambridge, Massachusetts, USA, March 24-26, 2008.

[1,18] and on lattice problems over rings such as NTRU [21], PKCs based on braid groups [25] and PKCs based on Multivariate Quadratic (MQ) polynomials.

The main *spiritus movens* for proposing new PKCs is the following wide-spread cryptographic folklore wisdom: *"it is not wise to keep all the eggs in one basket"*. However, from the performance point of view, there is always another practical motivation: *"to design secure public key cryptosystems that has better speed performance"*.

### 1.1 Public key schemes based on MQ polynomials

There are 4 basic trapdoor functions that are based on multivariate quadratic polynomials. An excellent survey for those four classes of multivariate quadratic public key cryptosystems has been made by Wolf and Preneel in [47]. Here we give a brief summary.

The first MQ scheme called MIA is that of Matsumoto and Imai [22] from 1985. That scheme was broken in 1995 by Patarin [33].

The second MQ scheme called STS (Stepwise Triangular Scheme) was first introduced in 1993 by Shamir [41] in the variant called Birational Permutation Schemes and was successfully broken by Coppersmith, Stern and Vaudenay in the same year [6]. In 1999 a MQ scheme which is a variant of STS called TTM was proposed by Moh [32]. That scheme was broken in 2000 by Goubin and Courtois [19]. They generalized the STS into the scheme TPM (Triangle Plus Minus) for which the TTM of Moh is a special case. The whole STS class has been broken by Wolf, Braeken and Preneel in 2004 [44].

The third MQ scheme called HFE (Hidden Field Equations) was designed by Patarin [34,35] in 1996. It is a sort of generalization of the MIA scheme. Basic HFE was broken by solving instances of the MinRank problem, by Kipnis and Shamir [24] in 1999. Several modifications of the scheme have been proposed [45] but also several new successful attacks have been published [46,48].

The forth scheme called UOV (Unbalanced Oil and Vinegar) was proposed in 1999 by Kipnis, Patarin, and Goubin [23] and is a generalization of the original Oil and Vinegar scheme of Patarin [36] from 1997. Some basic variants of UOV have been successfully broken in [4,48]. The main design flaw of this scheme is the fact that one set of variables (vinegar variables) are combined quadratically, but the other complementary set of variables (oil variables) are combined with only vinegar variables in a quadratic way. This design characteristic was the source of some successful attacks on this scheme. Also, this algorithm can only be used for signature schemes. However, for some carefully chosen parameters the scheme is considered still not broken.

Wolf and Preneel [47] have also indicated numerous ways how to tweak all those MQ systems.

### 1.2 Our results

We have designed a new class of MQ trapdoor functions. The generation of our trapdoor functions is based on the theory of quasigroups and quasigroup string transformations. In Section 2 we give a brief introduction to quasigroups and quasigroup string transformations. In the same section we define a new special class of so called *Multivariate Quadratic Quasigroups (MQQ)*. In Section 3 we describe a public-key cryptosystem based on MQQs. Its operating characteristics are given in Section 4. We discuss the security of our PKC in Section 5. Conclusions are given in Section 6.

The results about our PKC can be briefly summarized as:

- it is a deterministic one-to-one mapping;
- there is no message expansion;
- it has one parameter $n$ $(= 140, 160, 180, 200 \ldots)$ – the bit length of the encrypted block;
- its conjectured security level when $n \geq 140$ bits is $2^{\frac{n}{2}}$;
- its encryption speed is comparable to the speed of other multivariate quadratic PKCs;
- its decryption/signature speed is as a typical symmetric block cipher (i.e., in the range of 500–1000 times faster than the most popular public key schemes);
- it is well suited for short signatures.

## 2 Preliminaries

In this section we will briefly introduce quasigroup string transformations in 2.1, representation of the quasigroups as vector valued Boolean functions in 2.2, we will discuss the class of multivariate quadratic quasigroups in 2.3, and the bijection of Dobbertin in 2.4.

### 2.1 Quasigroup string transformations

Here we give a brief overview of quasigroups and quasigroup string transformations. A more detailed explanation is found in [3,8,28,29,42].

**Definition 1.** *A quasigroup $(Q, *)$ is a groupoid satisfying the law*

$$(\forall u, v \in Q)(\exists! x, y \in Q) \quad u * x = v \ \& \ y * u = v. \tag{1}$$

It follows from (1) that for each $a, b \in Q$ there is a unique $x \in Q$ such that $a * x = b$. Then we denote $x = a \setminus_* b$ where $\setminus_*$ is a binary operation in $Q$ (called a left parastrophe of $*$) and the groupoid $(Q, \setminus_*)$ is a quasigroup too. The algebra $(Q, *, \setminus_*)$ satisfies the identities

$$x \setminus_* (x * y) = y, \quad x * (x \setminus_* y) = y. \tag{2}$$

Consider an alphabet (i.e., a finite set) $Q$, and denote by $Q^+$ the set of all nonempty words (i.e., finite strings) formed by the elements of $Q$. In this paper, depending on the context, we will use two notifications for the elements of $Q^+$: $a_1 a_2 \dots a_n$ and $(a_1, a_2, \dots, a_n)$, where $a_i \in Q$. Let $*$ be a quasigroup operation on the set $Q$. For each $l \in Q$ we define two functions $e_{l,*}, d_{l,*} : Q^+ \to Q^+$ as follows:

**Definition 2.** *Let $a_i \in Q$, $M = a_1 a_2 \dots a_n$. Then*

$$e_{l,*}(M) = b_1 b_2 \dots b_n \iff$$
$$b_1 = l * a_1,\ b_2 = b_1 * a_2, \dots,\ b_n = b_{n-1} * a_n,$$
$$d_{l,*}(M) = c_1 c_2 \dots c_n \iff$$
$$c_1 = l * a_1,\ c_2 = a_1 * a_2, \dots,\ c_n = a_{n-1} * a_n,$$

*i.e., $b_{i+1} = b_i * a_{i+1}$ and $c_{i+1} = a_i * a_{i+1}$ for each $i = 0, 1, \dots, n-1$, where $b_0 = a_0 = l$.*

The functions $e_{l,*}$ and $d_{l,*}$ are called the $e$–transformation and the $d$–transformation of $Q^+$ based on the operation $*$ with leader $l$ respectively. Graphical representations of $e$–transformation and $d$–transformation are shown in Fig. 1.

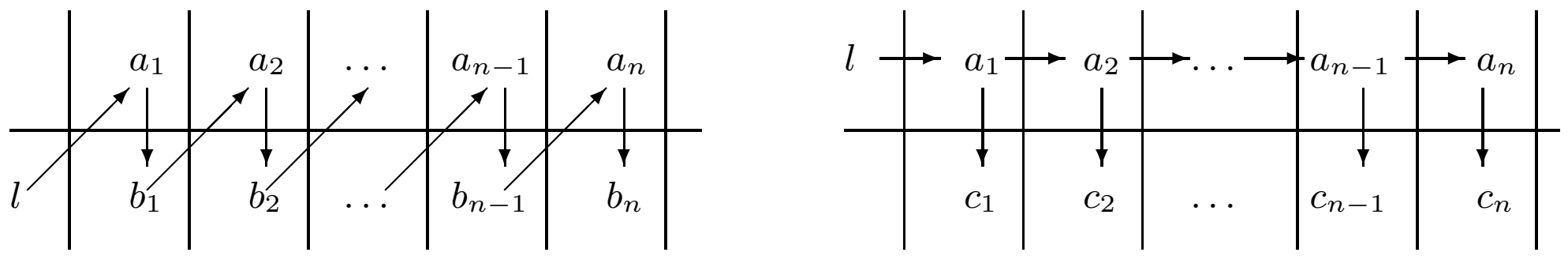

**Fig. 1.** Graphical representations of the $e_{l,*}$ and $d_{l,*}$ transformations

**Theorem 1.** *If $(Q, *)$ is a finite quasigroup, then $e_{l,*}$ and $d_{l,\setminus_*}$ are mutually inverse permutations of $Q^+$, i.e.,*

$$d_{l,\setminus_*}(e_{l,*}(M)) = M = e_{l,*}(d_{l,\setminus_*}(M))$$

*for each leader $l \in Q$ and for every string $M \in Q^+$.* ■

### 2.2 Quasigroups as vector valued Boolean functions

To define a multivariate quadratic PKC for our purpose, we will use the presentation of finite quasigroups $(Q, *)$ of order $2^d$ by vector valued Boolean functions (v.v.b.f.). Consequently, we choose a bijection $\beta : Q \to \{0, 1, \ldots, 2^d - 1\}$ and represent $a \in Q$ by the $d$-bit representation $\beta(a)$. Hence, for each $a \in Q$ there are uniquely determined bits $x_1, x_2, \ldots, x_d \in \{0, 1\}$ (which depend on the choice of the bijection $\beta$) such that $a$ is represented by the string $x_1x_2 \ldots x_d$. Then we identify $a$ and its $d$-bit representation and write $a = x_1x_2 \ldots x_d$ or, sometimes, $a = (x_1, x_2, \ldots, x_d)$. Now, the binary operation $*$ on $Q$ can be seen as a vector valued operation $*_{vv} : \{0, 1\}^{2d} \to \{0, 1\}^d$ defined as:

$$a * b = c \Longleftrightarrow$$
$$*_{vv}(x_1, x_2, \ldots, x_d, y_1, y_2, \ldots, y_d) = (z_1, z_2, \ldots, z_d),$$

where $x_1 \ldots x_d,\ y_1 \ldots y_d,\ z_1 \ldots z_d$ are binary representations of $a, b, c$ respectively.

Each $z_i$ depends of the bits $x_1, x_2, \ldots, x_d, y_1, y_2, \ldots, y_d$ and is uniquely determined by them. So, each $z_i$ can be seen as a $2d$-ary Boolean function $z_i = f_i(x_1, x_2, \ldots, x_d, y_1, y_2, \ldots, y_d)$, where $f_i : \{0, 1\}^{2d} \to \{0, 1\}$ strictly depends on, and is uniquely determined by, $*$. Thus, we have the following:

**Lemma 1.** *For every quasigroup $(Q, *)$ of order $2^d$ and for each bijection $Q \to \{0, 1 \ldots, 2^d - 1\}$ there are a uniquely determined v.v.b.f. $*_{vv}$ and $d$ uniquely determined $2d$-ary Boolean functions $f_1, f_2, \ldots, f_d$ such that for each $a, b, c \in Q$*

$$a * b = c \Longleftrightarrow *_{vv}(x_1, \ldots, x_d, y_1, \ldots, y_d) =$$
$$= (f_1(x_1, \ldots, x_d, y_1, \ldots, y_d), ..., f_d(x_1, \ldots, x_d, y_1, \ldots, y_d)). \qquad \blacksquare$$

Recall that each $k$-ary Boolean function $f(x_1, \ldots, x_k)$ can be represented in a unique way by its algebraic normal form (ANF), i.e., as a sum of products

$$ANF(f) = \alpha_0 + \sum_{i=1}^{k} \alpha_i x_i + \sum_{1 \le i < j \le k} \alpha_{i,j} x_i x_j + \sum_{1 \le i < j < s \le k} \alpha_{i,j,s} x_i x_j x_s + \ldots, \tag{3}$$

where the coefficients $\alpha_0, \alpha_i, \alpha_{i,j}, \ldots$ are in the set $\{0, 1\}$ and the addition and multiplication are in the field $GF(2)$. In the rest of the text we will abuse the notation and identify the Boolean function $f$ and its ANF, i.e., we will take $f = ANF(f)$. We say a polynomial $f(x_1, \ldots, x_k)$ when we consider the arguments of $f$ to be indeterminate variables $x_1, x_2, \ldots, x_k$.

The ANFs of the functions $f_i$ give us information about the complexity of the quasigroup $(Q, *)$ via the degrees of the Boolean functions $f_i$. It can be observed that the degrees of the polynomials $ANF(f_i)$ rise with the order of the quasigroup. In general, for a randomly generated quasigroup of order $2^d$, $d \ge 4$, the degrees are higher than 2. Such quasigroups are not suitable for our construction of multivariate quadratic PKC.

### 2.3 Multivariate Quadratic Quasigroups

In this subsection we define a special class of quasigroups, called *multivariate quadratic quasigroups* (MQQs) that can be of different types.

**Definition 3.** *A quasigroup $(Q, *)$ of order $2^d$ is called Multivariate Quadratic Quasigroup (MQQ) of type $Quad_{d-k}Lin_k$ if exactly $d - k$ of the polynomials $f_i$ are of degree 2 (i.e., are quadratic) and $k$ of them are of degree 1 (i.e., are linear), where $0 \le k < d$.*

Theorem 2 below gives us sufficient conditions for a quasigroup $(Q, *)$ to be MQQ.

**Theorem 2.** *Let $\mathbf{A_1} = [f_{ij}]_{d \times d}$ and $\mathbf{A_2} = [g_{ij}]_{d \times d}$ be two $d \times d$ matrices of linear Boolean expressions, and let $\mathbf{b_1} = [u_i]_{d \times 1}$ and $\mathbf{b_2} = [v_i]_{d \times 1}$ be two $d \times 1$ vectors of linear or quadratic Boolean expressions.*

*Let the functions $f_{ij}$ and $u_i$ depend only on variables $x_1, \ldots, x_d$, and let the functions $g_{ij}$ and $v_i$ depend only on variables $x_{d+1}, \ldots, x_{2d}$. If*

$$\mathbf{Det}(\mathbf{A_1}) = \mathbf{Det}(\mathbf{A_2}) = 1 \textit{ in } GF(2) \tag{4}$$

*and if*

$$\mathbf{A_1} \cdot (x_{d+1}, \ldots, x_{2d})^T + \mathbf{b_1} \equiv \mathbf{A_2} \cdot (x_1, \ldots, x_d)^T + \mathbf{b_2} \tag{5}$$

*then the vector valued operation $*_{vv}(x_1, \ldots, x_{2d}) = \mathbf{A_1} \cdot (x_{d+1}, \ldots, x_{2d})^T + \mathbf{b_1}$ defines a quasigroup $(Q, *)$ of order $2^d$ that is MQQ.*

*Proof.* Consider the equation

$$*_{vv}(a_1, \ldots, a_d, x_{d+1}, \ldots, x_{2d}) = (c_1, \ldots, c_d)$$

where $x_{d+1}, \ldots, x_{2d}$ are unknown bits, while $a_i$, $c_i$ are given bits. We have the linear system in $GF(2)$ of kind

$$\mathbf{A'_1} \cdot (x_{d+1}, \ldots, x_{2d})^T + \mathbf{b'_1} = (c_1, \ldots, c_d)^T, \tag{6}$$

where $\mathbf{A'_1}$ and $\mathbf{b'_1}$ are the valuations of $\mathbf{A_1}$ and $\mathbf{b_1}$ over the vector $(a_1, \ldots, a_d)$. Since $\mathbf{Det}(\mathbf{A_1}) = 1$, it follows that $\mathbf{Det}(\mathbf{A'_1}) = 1$ too, so the linear system (6) has a unique solution $(x_{d+1}, \ldots, x_{2d})^T = (\mathbf{A'_1})^{-1} \cdot ((c_1, \ldots, c_d)^T - \mathbf{b'_1})$. In the same manner a unique solution of the equation

$$*_{vv}(x_1, \ldots, x_d, a_{d+1}, \ldots, a_{2d}) = (c_1, \ldots, c_d)$$

can be found, and $*_{vv}$ is a v.v.b.f. of a quasigroup operation $*$ on the set $Q = \{0, 1, \ldots, 2^d - 1\}$. The quasigroup $(Q, *)$ is MQQ since the vector $\mathbf{A_1} \cdot (x_{d+1}, \ldots, x_{2d})^T$ has as elements multivariate quadratic polynomials.

*Example 1.* Let the quasigroup $(Q, *)$ of order $2^3 = 8$ be given by the multiplication scheme in Table 1.

| * | 0 | 1 | 2 | 3 | 4 | 5 | 6 | 7 |
|---|---|---|---|---|---|---|---|---|
| 0 | 3 | 2 | 6 | 7 | 1 | 0 | 4 | 5 |
| 1 | 5 | 3 | 7 | 1 | 0 | 6 | 2 | 4 |
| 2 | 0 | 6 | 3 | 5 | 4 | 2 | 7 | 1 |
| 3 | 6 | 7 | 2 | 3 | 5 | 4 | 1 | 0 |
| 4 | 7 | 1 | 4 | 2 | 3 | 5 | 0 | 6 |
| 5 | 1 | 0 | 5 | 4 | 2 | 3 | 6 | 7 |
| 6 | 4 | 5 | 1 | 0 | 6 | 7 | 3 | 2 |
| 7 | 2 | 4 | 0 | 6 | 7 | 1 | 5 | 3 |

| \ | 0 | 1 | 2 | 3 | 4 | 5 | 6 | 7 |
|---|---|---|---|---|---|---|---|---|
| 0 | 5 | 4 | 1 | 0 | 6 | 7 | 2 | 3 |
| 1 | 4 | 3 | 6 | 1 | 7 | 0 | 5 | 2 |
| 2 | 0 | 7 | 5 | 2 | 4 | 3 | 1 | 6 |
| 3 | 7 | 6 | 2 | 3 | 5 | 4 | 0 | 1 |
| 4 | 6 | 1 | 3 | 4 | 2 | 5 | 7 | 0 |
| 5 | 1 | 0 | 4 | 5 | 3 | 2 | 6 | 7 |
| 6 | 3 | 2 | 7 | 6 | 0 | 1 | 4 | 5 |
| 7 | 2 | 5 | 0 | 7 | 1 | 6 | 3 | 4 |

**Table 1.** A quasigroup $(Q, *)$ and its left parastrophe $(Q, \backslash)$ of order 8.

The corresponding ANF representation of the operation $*$ as a vector valued Boolean function is the following:

$$*_{vv}(x_1, x_2, x_3, x_4, x_5, x_6) = (f_1, f_2, f_3),$$

where
$f_1 = x_1 + x_3 + x_1x_4 + x_2x_4 + x_3x_4 + x_5 + x_1x_5 + x_2x_5 + x_3x_5 + x_1x_6 + x_2x_6 + x_3x_6,$
$f_2 = 1 + x_2 + x_3 + x_4 + x_1x_4 + x_2x_4 + x_3x_4 + x_1 + x_5 + x_2x_5 + x_3x_5 + x_1x_6 + x_2x_6 + x_3x_6,$
$f_3 = 1 + x_2 + x_3x_4 + x_5 + x_3x_5 + x_6 + x_1x_6 + x_2x_6 + x_3x_6.$

The corresponding matrix-vector representations of $*$ by $\mathbf{A_1}$, $\mathbf{b_1}$, and $\mathbf{A_2}$,$\mathbf{b_2}$, are the following:

$$*_{vv}(x_1, x_2, x_3, x_4, x_5, x_6) = \mathbf{A_1}(x_4, x_5, x_6)^T + \mathbf{b_1},$$

where $\mathbf{A_1} = \begin{bmatrix} x_1+x_2+x_3 & 1+x_1+x_2+x_3 & x_1+x_2+x_3 \\ 1+x_1+x_2+x_3 & x_1+x_2+x_3 & x_1+x_2+x_3 \\ x_3 & 1+x_3 & 1+x_1+x_2+x_3 \end{bmatrix}$ and $\mathbf{b_1} = \begin{bmatrix} x_1+x_3 \\ 1+x_2+x_3 \\ 1+x_2 \end{bmatrix}$;

$$*_{vv}(x_1,x_2,x_3,x_4,x_5,x_6) = \mathbf{A_2}(x_1,x_2,x_3)^T + \mathbf{b_2},$$

where $\mathbf{A_2} = \begin{bmatrix} 1+x_4+x_5+x_6 & x_4+x_5+x_6 & 1+x_4+x_5+x_6 \\ x_4+x_5+x_6 & 1+x_4+x_5+x_6 & 1+x_4+x_5+x_6 \\ x_6 & 1+x_6 & x_4+x_5+x_6 \end{bmatrix}$ and $\mathbf{b_2} = \begin{bmatrix} x_5 \\ 1+x_4 \\ 1+x_5+x_6 \end{bmatrix}$.

Indeed in $GF(2)$, $\mathbf{Det}(\mathbf{A_1}) = \mathbf{Det}(\mathbf{A_2}) = 1$.

While in this example the quasigroup $(Q,*)$ is multivariate quadratic, its corresponding left parastrophe $(Q,\backslash)$ is of higher degree (the degree is 3). Actually it is a typical behavior for the set of all MQQs. Their corresponding left parastrophes have ANF representation that has higher degree than 2. The ANF representation for $(Q,\backslash)$ in this example is the following:

$$\backslash_{vv}(x_1,x_2,x_3,x_4,x_5,x_6) = (g_1,g_2,g_3)$$

where

$g_1 = 1+x_2+x_1x_3+x_2x_3+x_1x_4+x_2x_4+x_1x_3x_4+x_2x_3x_4+x_5+x_3x_5+x_1x_3x_5+x_2x_3x_5+x_1x_6+x_2x_6+x_3x_6$,

$g_2 = x_1+x_1x_3+x_2x_3+x_4+x_1x_4+x_2x_4+x_1x_3x_4+x_2x_3x_4+x_3x_5+x_1x_3x_5+x_2x_3x_5+x_1x_6+x_2x_6+x_3x_6$,

$g_3 = 1+x_1+x_2+x_3+x_4+x_1x_4+x_2x_4+x_1x_5+x_2x_5+x_6$.

By using Theorem 2 we define the procedure MQQ$(d,k)$ for producing MQQs of order $2^d$ and of type $Quad_{d-k}Lin_k$ (see the Table 2).

| **MQQ**$(d,k)$ |
|---|
| **Input:** Integer $d$ and integer $k$, $0 \le k < d$ |
| **Output:** a quasigroup of order $2^d$ and of type $Quad_{d-k}Lin_k$ |
| 1. Randomly generate a $d \times d$ matrix $\mathbf{A_1}$ of linear Boolean expressions of variables $x_1,\dots,x_d$, such that $\mathbf{Det}(\mathbf{A_1}) = 1$ in $GF(2)$ and the number $\#Const$ of constants 0 or 1 in the matrix $\mathbf{A_1}$ satisfies the inequality $kd \le \#Const < (k+1)d$ .<br>2. Randomly generate a $d \times 1$ vector $\mathbf{b_1}$ of linear Boolean expressions of variables $x_1,\dots,x_d$.<br>3. Compute the vector $*_{vv} = \mathbf{A_1}\cdot\mathbf{x_2} + \mathbf{b_1}$, where $\mathbf{x_2} = (x_{d+1},\dots,x_{2d})^T$.<br>4. Represent $*_{vv}$ as $*_{vv} = \mathbf{A_2}\cdot\mathbf{x_1} + \mathbf{b_2}$, where $\mathbf{x_1} = (x_1,\dots,x_d)^T$.<br>5. **if** ($\mathbf{Det}(\mathbf{A_2}) = 1$ in $GF(2)$) **and** ($*_{vv}$ is of type $Quad_{d-k}Lin_k$)<br>**then** Return($*_{vv}$),<br>**else GoTo 1;** |

**Table 2.** Heuristic algorithm for finding MQQs of order $2^5$

Note that the procedure MQQ$(d,k)$ is a randomized algorithm for finding MQQs of order $2^d$ and of type $Quad_{d-k}Lin_k$. For $d=5$ the average number of attempts for finding MQQs of type $Quad_4Lin_1$ is around $2^{15}$ and for finding MQQs of type $Quad_5Lin_0$ is around $2^{16}$. However, MQQ$(6,0)$ did not give us any MQQ of order $2^6$. Finding MQQs of orders $2^d, d \ge 6$, we consider as an open research problem.

The definition of MQQs implies the following theorem:

**Theorem 3.** *Let* $\mathbf{x_1} = (f_1,f_2,\dots,f_d)$ *and* $\mathbf{x_2} = (f_{d+1},f_{d+2},\dots,f_{2d})$ *be two d–dimensional vectors of linear Boolean functions of variables* $x_1,\dots,x_d$. *Let* $(Q,*)$ *be a multivariate quadratic quasigroup of type* $Quad_{d-k}Lin_k$. *If* $\mathbf{x_1} * \mathbf{x_2} = (g_1,\dots,g_d)$ *then at most* $d-k$ *of the polynomials* $g_i$ *are multivariate quadratic and at least* $k$ *polynomials are linear.* ∎

We want to emphasize that in a process of a random generation of MQQs of type $Quad_{d-k}Lin_k$, usually the number of quadratic polynomials is exactly $d-k$, and the number of linear polynomials is exactly $k$. However, there are rare cases when all quadratic terms can cancel each other, and the number of linear polynomials will be bigger than $k$ while the number of quadratic polynomials will be less than $d-k$. Nevertheless, these cases, if they occur, can be easily detected, and quasigroups with such properties can be omitted from consideration as a candidates for the private key.

### 2.4 The bijection of Dobbertin

In our construction of the multivariate quadratic trapdoor bijective functions we mostly use the quasigroups defined in the previous subsections with properties given in Theorem 1 and Theorem 3. However, by using only MQQs some of the coordinate functions will remain linear. In order to make a trapdoor bijective function $\{0,1\}^n \to \{0,1\}^n$ that is multivariate quadratic in all of its coordinates we use the bijection of Dobbertin.

Dobbertin has proved [11] that the function $\mathbf{Dob}(X) = X^{2^{m+1}+1} + X^3 + X$ is a bijection in $GF(2^{2m+1})$. Moreover it is multivariate quadratic too.

In our design of MQQ public key cryptosystem we use the bijection of Dobbertin for $m = 6$.

## 3 Description of the algorithm

A generic description for our scheme can be expressed as a typical multivariate quadratic system: $T \circ P' \circ S : \{0,1\}^n \to \{0,1\}^n$ where $T$ and $S$ are two nonsingular linear transformations, and $P'$ is a bijective multivariate quadratic mapping on $\{0,1\}^n$.

First we will describe how the mapping $P' : \{0,1\}^n \to \{0,1\}^n$ is defined by the algorithm described in Table 3.

For the reasons why in the "Preprocessing phase" we demand that "minimal rank of their quadratic polynomials when represented in matrix form is at least 8" please see the Subsection 5.4.

Additionally, note that the definition of the index set $I = (i_1, i_2, \ldots, i_{k-1})$ where $i_j \in \{1, 2, \ldots, 8\}$ can be either public or private. The security of the algorithm does not depend on the secrecy of that set.

The algorithm for generating the public and private key is defined in the Table 4. We give in the Appendix a detailed example of the process of generating the public and private key with small number of variables $n = 20$.

The algorithm for decryption/signing by the private key $(T, S, *_1, \ldots, *_8)$ is defined in Table 5.

The algorithm for encryption with the public key is straightforward application of the set of $n$ multivariate polynomials $\mathbf{P} = \{P_i(x_1, \ldots, x_n) \mid i = 1, \ldots, n\}$ over a vector $x = (x_1, \ldots, x_n)$, i.e., $y = \mathbf{P}(x)$.

## 4 Operating characteristics

In this section we will discuss the size of the private and the public key as well as the number of operations per byte for encryption and decryption.

### 4.1 The size of the public and the private key

Since the public key consists of $n$ multivariate quadratic equations, and they appear to be randomly generated, the size of the public key follows the rules given in [47]. So, for $n$ bit blocks the size of the public key is $n \times (1 + \frac{n(n+1)}{2})$ bits. In the Table 6 we give the size of the public key for $n \in \{140, 160, 180, 200\}$ in KBytes.

The private key of our scheme is the tuple $(T, P, *_1, \ldots, *_8)$. The corresponding memory size needed for storage of $T$ and $P$ is $2n^2$ bits. The memory size for the quasigroups $(*_1, \ldots, *_8)$, (actually for their parastrophes), is $8 \times 32 \times 32 \times 5 = 40960$ bits. For the storage of particular quasigroups in memory we note that it is not necessary to store $32 \times 32 \times 5$ bits for every quasigroup, since that type of the storage has redundancy (the last row of the Latin Square is uniquely determined by the rest of the table), but in order to achieve efficient speed in the decryption process we store the full information about the parastrophes. In total, the size of the private key expressed in Kb is $\frac{1}{2^{13}}(2n^2 + 40960)$. In the second column of the Table 6 we give the size of the private key for $n \in \{140, 160, 180, 200\}$ in KB (kilo bytes).

| $P'(n)$ |
|---|
| **Input:** Integer $n$, where $n = 5k$, $k \geq 28$, and a vector $x = (f_1, \ldots, f_n)$ of $n$ linear Boolean functions of $n$ variables. |
| **Output:** Eight quasigroups $*_1, \ldots, *_8$ and $n$ multivariate quadratic polynomials $P'_i(x_1, \ldots, x_n), i = 1, \ldots, n$ |
| **Preprocessing phase**<br>By calling the procedure MQQ(4,1) and MQQ(5,0) generate two large sets $\mathbf{Quad_4Lin_1}$ and $\mathbf{Quad_5Lin_0}$ (with more than $2^{20}$ elements each) of MQQs of type $Quad_4Lin_1$ and of type $Quad_5Lin_0$ such that the minimal rank of their quadratic polynomials when represented in matrix form is at least 8;<br>Transform (by permuting the coordinates) all quasigroups in the set $\mathbf{Quad_4Lin_1}$ such that their first coordinate is linear. |
| 1. Represent a vector $x = (f_1, \ldots, f_n)$ of $n$ linear Boolean functions of $n$ variables $x_1, \ldots, x_n$, as a string $x = X_1 \ldots X_k$ where $X_i$ are vectors of dimension 5;<br>2. Pick randomly different quasigroups $*_1, *_2 \in \mathbf{Quad_4Lin_1}$ and different quasigroups $*_3, *_4, *_5, *_6, *_7, *_8 \in \mathbf{Quad_5Lin_0}$.<br>3. Define a $(k-1)$–tuple $I = (i_1, i_2, \ldots, i_{k-1})$ where $i_j \in \{1, 2, \ldots, 8\}$, that will used as an index set (sequence) to determine which quasigroup will be used in the nonlinear transformation of $y$. The requirement for this index set is that the total number of indexes that are refereing to a quasigroup from the class $\mathbf{Quad_4Lin_1}$ is 8.<br>4. Compute $y = Y_1 \ldots Y_k$ where: $Y_1 = X_1$, $Y_{j+1} = X_j *_{i_j} X_{j+1}$, for $j = 1, 2, \ldots, k-1$<br>5. Set a 13-dimensional vector $Z = Y_1 \vert\vert Y_{\mu_1,1} \vert\vert Y_{\mu_2,1} \vert\vert \ldots \vert\vert Y_{\mu_8,1}$ that has all 13 components as linear Boolean functions. Here the notation $Y_{\mu_j,1}$ means the first coordinate of the vector $Y_{\mu_j}$.<br>6. Transform $Z$ by the bijection of Dobbertin: $W = \mathbf{Dob}(Z)$.<br>7. Set $Y_1 = (W_1, W_2, W_3, W_4, W_5)$, $Y_{\mu_1,1} = W_6$, $Y_{\mu_2,1} = W_7$, $Y_{\mu_3,1} = W_8$, $Y_{\mu_4,1} = W_9$, $Y_{\mu_5,1} = W_{10}$, $Y_{\mu_6,1} = W_{11}$, $Y_{\mu_7,1} = W_{12}$, $Y_{\mu_8,1} = W_{13}$.<br>8. Output: Quasigroups $*_1, \ldots, *_8$ and $y$ as $n$ multivariate quadratic polynomials $P'_i(x_1, \ldots, x_n),\ i = 1, \ldots, n$. |

**Table 3.** Definition of the nonlinear mapping $P' : \{0,1\}^n \to \{0,1\}^n$

| Algorithm for generating Public and Private key for the MQQ scheme |
|---|
| **Input:** Integer $n$, where $n = 5k$ and $k \geq 28$. |
| **Output:** Public key $\mathbf{P}$: $n$ multivariate quadratic polynomials $P_i(x_1, \ldots, x_n),\ i = 1, \ldots, n$,<br>Private key: Two nonsingular Boolean matrices $T$ and $S$ of order $n \times n$ and eight quasigroups $*_1, \ldots, *_8$ |
| 1. Generate two nonsingular $n \times n$ Boolean matrices $T$ and $S$ (uniformly at random).<br>2. Call the procedure for definition of $P'(n) : \{0,1\}^n \to \{0,1\}^n$ and from there also obtain the quasigroups $*_1, \ldots, *_8$.<br>3. Compute $y = T(P'(S(x)))$ where $x = (x_1, \ldots, x_n)$.<br>4. Output: The public key is $y$ as $n$ multivariate quadratic polynomials $P_i(x_1, \ldots, x_n), i = 1, \ldots, n$, and the private key is the tuple $(T, S, *_1, \ldots, *_8)$. |

**Table 4.** Algorithm for generating the public and private key

For the storage of the inverse table of the bijection of Dobbertin we need additional $2^{13} \times 13 = 106496$ bits which is exactly 13KB, but those 13KB do not belong to the private key.

### 4.2 The number of operations for encryption and decryption

In order to obtain an independent measure for the operating speed of our scheme, we will express the speed of encryption and decryption/signing as the number of operations per processed byte. We will also take into account three widespread microprocessor architectures: 8-bit, 32-bit and 64-bit architectures.

| Algorithm for decryption/signing with the private key $(T, S, *_1, \ldots, *_8)$ |
|---|
| **Input:** A vector $y = (y_1, \ldots, y_n)$. |
| **Output:** A vector $x = (x_1, \ldots, x_n)$ such that $\mathbf{P}(x) = y$. |
| 1. Set $y' = T^{-1}(y)$.<br>2. Set $W = (y'_1,\ y'_2,\ y'_3,\ y'_4,\ y'_5,\ y'_6,\ y'_{11},\ y'_{16},\ y'_{21},\ y'_{26},\ y'_{31},\ y'_{36},\ y'_{41})$.<br>3. Compute $Z = (Z_1, Z_2, Z_3, Z_4, Z_5, Z_6, Z_7, Z_8, Z_9, Z_{10}, Z_{11},\ Z_{12}, Z_{13}) = \mathbf{Dob}^{-1}(W)$.<br>4. Set $y'_1 \leftarrow Z_1,\ y'_2 \leftarrow Z_2,\ y'_3 \leftarrow Z_3,\ y'_4 \leftarrow Z_4,\ y'_5 \leftarrow Z_5,\ y'_6 \leftarrow Z_6,\ y'_{11} \leftarrow Z_7\ y'_{16} \leftarrow Z_8,\ y'_{21} \leftarrow Z_9,$ $y'_{26} \leftarrow Z_{10},\ y'_{31} \leftarrow Z_{11}\ y'_{36} \leftarrow Z_{12},\ y'_{41} \leftarrow Z_{13}$.<br>5. Represent $y'$ as $y' = Y_1 \ldots Y_k$ where $Y_i$ are vectors of dimension 5.<br>6. By using the left parastrophes $\backslash_i$ of the quasigroups $*_i$, $i = 1, \ldots, 8$, obtain $x' = X_1 \ldots X_k$, such that: $X_1 = Y_1$, $X_2 = X_1 \backslash_1 Y_2$, $X_3 = X_2 \backslash_2 Y_3$ and $X_i = X_{i-1} \backslash_{3+((i+2) \bmod 6)} Y_i$.<br>7. Compute $x = S^{-1}(x')$. |

**Table 5.** Algorithm for decryption or signing

| $n$ | Size of the public key (KBytes) | Size of the private key (KBytes) |
|---|---|---|
| 140 | 168.69 | 9.79 |
| 160 | 251.58 | 11.25 |
| 180 | 357.96 | 12.91 |
| 200 | 490.75 | 14.77 |

**Table 6.** Memory size in KBytes for the public key and the private key

Since the public part of our scheme follows the typical paradigm of the MQ public key cryptosystems, its speed of encryption is the same as (or similar to) the speed of other MQ systems. That means that the encryption is done after $O(n^3)$ logical AND and logical XOR operations.

The actual speed of any multivariate quadratic PKC when encryption is performed on 32-bit or 64-bit microprocessor architectures, using internal parallelism of the modern CPUs, as well as techniques of bit slicing, can result in an encryption process which is significantly faster than RSA/DH or ECC encryption for systems with equivalent security levels.

If we assume that AND or XOR operations can be executed in one cycle (without taking into account that modern 32-bit and 64-bit CPUs actually can perform several such operations in parallel), then non-optimized encryption of any general $n$-bit variant of any multivariate quadratic PKC scheme have a speed of $\frac{16}{n}\lceil \frac{n}{Arch} \rceil (1 + \frac{n(n+1)}{2})$ operations per byte where $Arch = 8,\ 32$ or 64.

The speed of decryption/signing in the class of multivariate quadratic PKCs is not so uniformly distributed as it is for encryption. The number of operations for particular parts of the process of decryption of our scheme can be summarized in the following list:

- Two linear operation by the matrices $S^{-1}$ and $T^{-1}$ that takes $2n\lceil \frac{n}{Arch} \rceil$ operations;
- One lookup operation at the table of the bijection of Dobbertin;
- Exactly $k - 1$ lookup operations at the quasigroup parastrophes.

The total number of operations per byte can be computed by the expression $\frac{8}{n}(2n\lceil \frac{n}{Arch} \rceil + 1 + k - 1)$ and are given in the Table 7.

### 4.3 Performance of the software and hardware implementation of the MQQ algorithm

We have implemented the MQQ public key algorithm in C and in VHDL, and have measured its performance both on a PC with Intel Core 2 Duo processor in 64–bit mode of operation, as well as on a Xilinx FPGA.

For comparative purposes, in Table 8 we give the speed of encryption/verification and decryption/signing for a basic block of data specific for several public key algorithms, performed on a PC with

| | Operations per encrypted byte | | |
|---|---|---|---|
| $n$ | Arch = 8-bit | Arch = 32-bit | Arch = 64-bit |
| 140 | 20306 | 5640 | 3384 |
| 160 | 25762 | 6440 | 3864 |
| 180 | 33306 | 8688 | 4344 |
| 200 | 40202 | 11257 | 6432 |

| | Operations per decrypted byte | | |
|---|---|---|---|
| $n$ | Arch = 8-bit | Arch = 32-bit | Arch = 64-bit |
| 140 | 289.65 | 81.66 | 49.66 |
| 160 | 321.65 | 81.65 | 49.65 |
| 180 | 369.64 | 97.64 | 49.64 |
| 200 | 401.64 | 113.64 | 65.64 |

**Table 7.** Estimated operations per encrypted/decrypted byte, for different $n$ and 8, 32 or 64 bit architectures.

Intel Core 2 Duo processor. The measurements for DSA, RSA, and ECDSA are taken from "eBATS: ECRYPT Benchmarking of Asymmetric Systems"[13]. From Table 8 one can obtain an impression that MQQ with 160–bits blocks is faster only in decryption (signatures), but is even slower than 1024–bit RSA in encryption (verification). However, we must emphasize the property that MQQ is highly parallelizable, and on systems with multiple CPU cores the speedup can go linearly with the number of CPU cores. That is shown in the last row of the Table 8, where 160–bit MQQ was implemented in C, using OpenMP 2.0 under Microsoft Visual Studio 2008. It is clear that further increasing of the number of CPU cores can speed up MQQ algorithm almost linearly with the number of cores. The next challenging task will be to implement MQQ in modern graphic cards (like Tesla™ many-core processor from NVIDIA).

| Algorithm name | Encrypt (cycles) | Decrypt (cycles) | Sign (cycles) | Verify (cycles) |
|---|---|---|---|---|
| DSA signatures using a 1024-bit prime | N/A | N/A | 1,041,400 | 1,246,312 |
| ECDSA signatures using NIST B-163 elliptic curve | N/A | N/A | 2,147,128 | 4,220,480 |
| 1024-bit RSA, 17 bits public exponent | 119,800 | 2,952,752 | 2,938,632 | 98,712 |
| 160–bit MQQ, one processor | 140,485 | 10,705 | 10,309 | 140,209 |
| 160–bit MQQ, two processors | 80,105 | 6,212 | 6,155 | 79,903 |

**Table 8.** Software speeds (in number of cycles) of several most popular public key algorithms on Intel Core 2 Duo processor in 64–bit mode of operation

The parallelizable nature of MQQ can be evidently shown in its hardware realization. Implemented in FPGA, MQQ is 10,000 times faster than DSA, RSA or ECDSA, and is comparable or even faster than the symmetric block cipher AES.

In Table 9 we compare the speed of 160–bit MQQ with the speed of 1024–bit RSA realized in Xilinx Virtex-5 FPGA chip by the company "Helion Technology Limited" [20]. In the same table we also give the speed of AES (128–bit key) realized in Xilinx FPGA chip in the same Virtex-5 family and by the same company. We have to mention that our FPGA realization of 160–bit MQQ is fully pipelined, and actually needs two Xilinx Virtex-5 chips of the type: XC5VFX70T-2.

| Algorithm name | 1024-bit RSA, encrypt/decrypt | 160–bit MQQ, encrypt/decrypt | 128–bit AES, encrypt/decrypt |
|---|---|---|---|
| FPGA type | Virtex-5, XC5VLX30-3 | Virtex-5, XC5VFX70T-2 | Virtex-5 |
| Frequency | 251 MHz | 276.7 / 249.4 MHz | 325 MHz |
| Throughput | 40 Kbps | 44.27 Gbps / 399.04 Mbps | 3.78 Gbps |

**Table 9.** Hardware performances of 1024–bit RSA, 160–bit MQQ and 128–bit AES on Xilinx Virtex-5 FPGAs

### 4.4 Speed of key generation for MQQ

In the Section 3 we have given a randomized algorithm for generating MQQs which is based on the heuristic algorithm for finding MQQs of order $2^5$ described in subsection 2.3. The whole process of public and private key is extremely time-consuming. So, from the speed of the key generation point of view, we can say that our MQQ algorithm has much worse characteristics compared to other public key algorithms.

## 5 Security analysis of the algorithm

### 5.1 The size of the pool of MQQs of order $2^5$

It is very important to address the question of the size of the set of MQQs of order $2^5$ and of types $Quad_5Lin_0$ and $Quad_4Lin_1$. We have to note that the number of MQQs of different types is an open research question and that we do not know the exact number of such quasigroups, nor an approximation for the lower bound for that number. In fact, a more thorough analysis and experiments with smaller quasigroups (of orders $2^2$ and $2^3$) can lead to the initial conjectures that their number is much larger than $2^{40}$. However, we are sure that their number is certainly larger than $2^{20}$ since, by using Mathematica as a package for symbolic computations on a modest Pentium 4 machine (2GHz, with 512MB RAM), we have generated more than $2^{20}$ MQQs of type $Quad_4Lin_1$ and more than $2^{20}$ MQQs of type $Quad_5Lin_0$. So, in total, the size of the pool set for 8 MQQs of order $2^5$ is at least of order $2^{149}$.

### 5.2 Chosen plaintext attack

The chosen plaintext attack against some MQ scheme first was successfully applied by Patarin in [33] to break the MIA scheme [22]. The idea is very simple: given a system of MQ polynomial equations $\{y_i = P_i(x_1, \ldots, x_n) \mid i = 1, \ldots, n\}$ as a public key, try to find relations in the $x$ and $y$ coordinates, by using plaintext/ciphertext pairs $(x, y)$.

It turned out that if the multivariate quadratic function is defined as $F : x \mapsto x^{q^l+1}$, then the relations between $x$ and $y$ coordinates can be expressed as a system of $n$ bilinear equations $\sum_{i=1}^{n} \sum_{j=1}^{n} \beta_{i,j} x_i y_j + \sum_{i=1}^{n} \beta_{i,0} x_i + \sum_{j=1}^{n} \beta_{0,j} y_j + \beta_{0,0} = 0$ where $\beta_{i,j}$ are $(n+1)^2$ unknown coefficients. By computing at least $(n+1)^2$ plaintext/ciphertext pairs $(x, y)$ a linear system of equations on unknowns $\beta_{i,j}$ can be established and solved.

Our MQ PKC scheme is not defined by any particular $F : x \mapsto x^{q^l+1}$. Moreover, if someone try to establish relations in the $x$ and $y$ coordinates in a similar way as Patarin did for MIA scheme, then those relations have to include all possible terms (not just bilinear terms) between $x$ and $y$ coordinates, thus rising the number of unknown terms to $O(2^n)$.

### 5.3 Isomorphism of polynomials

The isomorphism of polynomials with one secret was initially introduced by Patarin [34], and can be briefly formulated as: for two given sets of multivariate mappings $P' : \{0,1\}^n \to \{0,1\}^n$ and $P : \{0,1\}^n \to \{0,1\}^n$ by their polynomials $(P'_1(x_1,\dots,x_n),\dots,P'_n(x_1,\dots,x_n))$ and $(P_1(x_1,\dots,x_n), \dots, P_n(x_1,\dots,x_n))$ find (if any) an invertible affine mapping $S : \{0,1\}^n \to \{0,1\}^n$ such that $P' = S \circ P$, where the operation $\circ$ is a composition of mappings.

Several algorithms for the solution of the problem have been published in recent years. The first one was the algorithm proposed by Geiselmann, Meier and Steinwandt [17], and the second was the algorithm by Levy-dit-Vehel and Perret [27]. Finally, in 2005 Perret presented an algorithm for efficient solving of this problem which proved that the scheme of Patarin is not secure [38].

The main difficulty for applying the techniques developed by Perret for attacking our scheme is that our scheme differs from that of Patarin in the sense that it can be generally described as $P = T \circ P' \circ T \equiv T(P'(T(x)))$ where $T : \{0,1\}^n \to \{0,1\}^n$ is a nonsingular matrix and where the knowledge for $P'$ is hidden from the attacker. The bijection $P' : \{0,1\}^n \to \{0,1\}^n$ is constructed by an application of quasigroup transformations, and is not initially generated by some initial polynomial that will give information to the attacker.

### 5.4 The MinRank problem and its solutions as tools for attack

The MinRank problem can be described as follows: For a given sequence of matrices $(M_1,\dots,M_n)$ over some field and a given integer $r < n$, find a linear combination of the matrices such that $Rank(\sum_{i=1}^n \lambda_i M_i) \leq r$. The MinRank problem has been shown to be NP-complete over finite fields [5].

The attacks that involve solution of the MinRank problem, when $r$ is relatively small, work like this: represent a given public key of $n$ polynomials $P_1(x_1,\dots,x_n)$, …, $P_n(x_1,\dots,x_n)$ in the form $P_i(x_1,\dots,x_n) = x^T M_i x$ where $M_i$ are $n \times n$ matrices in a field. If the private key is constructed by polynomials $P'_i(x_1,\dots,x_n)$ such that they can be described as $P'_i(x_1,\dots,x_n) = x^T A_i x$, and if the minimal rank $r$ of $A_i$ is much smaller than $n$ ($r = 0, 1, 2$), then there are efficient algorithms for finding linear combination of the matrices $M_i$ such that $Rank(\sum_{i=1}^n \lambda_i M_i) \leq r$. The information for those linear combinations, combined with the particular design principles is then used to break the system. For example, Goubin and Courtois [19] have broken the TTM system of Moh [32] in this way.

The main difference between our PKC and systems where the solution of MinRank problems works is that the minimal rank $r$ of the matrices $A_i$ for the nonlinear part of our scheme are taken to be $r \geq 8$.

Kipnis and Shamir [24] have used an instance of the solution of MinRank problems to attack the HFE scheme. They used the fact that HFE polynomials can be described as $P(X) = \sum_{i=0}^{r-1}\sum_{j=0}^{r-1} p_{ij} X^{2^i+2^j}$, and the value of $r$ is kept small ($r \leq 13$) in order to implement efficient decryption.

This attack does not work on our scheme since we do not use any particular polynomial. In fact, if our nonlinear part $P'$ is represented as a polynomial then the values for $r$ are $\approx n$. Namely, the $P'$ is obtained by several applications of different nonlinear quasigroup string transformation, which are leaving some parts still linear. Then those linear leftovers are transformed by another nonlinear transformation. The resulting quadratic bijection $P' : \{0,1\}^n \to \{0,1\}^n$ can be represented in the form $P'(X) = \sum_{i=0}^{n-1}\sum_{j=0}^{n-1} p'_{ij} X^{2^i+2^j}$ only for $r \approx n$.

### 5.5 Attacks with differential cryptanalysis

A few years ago, Fouque, Granboulan and Stern [16] have come to the idea to use the concept of differential cryptanalysis, that has been used successfully mostly against symmetric cryptographic algorithms, for multivariate schemes too. The basic idea is that for any finite field $\mathbb{F}_q$ of characteristic $q$ and for any multivariate quadratic function $\mathbf{G} : (\mathbb{F}_q)^n \to (\mathbb{F}_q)^m$ the differential operator between any two points $x, k \in (\mathbb{F}_q)^n$ can be expressed as $\mathbf{L}_{\mathbf{G},k} = \mathbf{G}(x + k) - \mathbf{G}(x) - \mathbf{G}(k) + \mathbf{G}(0)$ and in fact that operator is a bilinear function.

By knowing the public key of a given multivariate quadratic scheme, and by knowing the information about the nonlinear part of that multivariate scheme (the function $F : x \mapsto x^{q^l+1}$), they showed that for certain parameters of some multivariate schemes it is possible to successfully recover the kernel of $\mathbf{L}_{\mathbf{G},k}$.

This attack was successfully applied on Ding's scheme [10], and afterwards, using the same technique, Dubois, Fouque, Shamir and Stern in [12] have completely broken all versions of the SFLASH signature scheme proposed by Patarin, Courtois, and Goubin in [37].

Although the generic part of the differential attack is applicable on our multivariate quadratic scheme, the crucial part that is different is the fact that the nonlinear part in our scheme is not a function of a type $F : x \mapsto x^{q^l+1}$. In our scheme, the nonlinear part is unknown to the attacker, since it is a part of the private key, and this renders the differential attack impoverished.

### 5.6 XL attack and Gröbner basis attacks

The XL [7] procedure for solving MQ polynomials was proposed by Courtois, Klimov, Patarin, and Shamir in 2000 as an extension of an initial attack by Kipnis and Shamir [24] called relinearization. When the number of equations $m$ is equal to the number of variables $n$, then the complexity of the algorithm is $2^n$. Since our PKC scheme has exactly that property, we can claim that an XL attack is not efficient on our PKC.

Recently XL algorithm has been proven ([2,43]) to be equivalent to F4 algorithm, a fast Gröbner basis computation algorithm developed by Faugere [14]. Since the Gröbner basis attack solves a multivariate polynomial system directly, any PKC based on multivariate polynomials is potentially vulnerable to this attack. In the case of HFE, Joux and Faugere [15] successfully attacked a system of 80 MQ polynomials in 80 variables in 2003 using Faugere's F5 algorithm specialized for the binary field. The successful breaking of HFE by F5 algorithm is due to the fact that the secret polynomial in the HFE has a degree of only 96. A HFE scheme need to have a polynomial of relatively low degree because it's performance depends on the algorithms for finding roots of that polynomial in $GF(2^n)$. The authors of [15] even give a table where they project the upper bound for the efficiency of the Gröbner basis attacks. The degree of the polynomials that can give HFE instances still feasible to be broken by the F5 algorithm goes up to 4096.

By performing numerous attacks by Faugere's algorithms on our PKC scheme we came to the following conclusions:

1. Faugere's algorithms successfully break our scheme up to $n = 95$ variables.
2. For $n \geq 100$, Faugere's algorithms rapidly lose their efficiency against our scheme.

We want to stress here that our decision to propose the security parameter for our PKC scheme (the number of Boolean variables $n$) to be: $n \geq 140$, was based both on the experimental experience that we gained during the *design-break-tweak-design* process and on the arguments that make our multivariate quadratic scheme different from all other such schemes developed so far. An important part in that design process were attacks based on Gröbner bases and Faugere's algorithms. However, as it is the case with many public-key schemes, we do not have an ultimate security proof for our scheme.

To summarize: as a general claim for our MQQ scheme with $n$ variables we say that its strength is $2^{\frac{n}{2}}$. We base our claims on the analysis of the power of the methods using Gröbner basis to solve random multivariate quadratic systems of equations.

Namely, the authors of [15] give a formula for computing the upper bound for the efficiency of the Gröbner basis attacks. Based on that analysis we are giving here the Table 10 with the projected complexity for solving random multivariate quadratic systems of equations by Gröbner basis algorithms for different number of variables $n$. Based on that projection in the second row we are giving the projection for the strength of our PKC scheme.

| $n$ | 140 | 160 | 180 | 200 |
|---|---|---|---|---|
| Complexity of Gröbner basis attacks | $2^{87}$ | $2^{99}$ | $2^{112}$ | $2^{125}$ |
| Strength of our MQQ PKC | $2^{70}$ | $2^{80}$ | $2^{90}$ | $2^{100}$ |

**Table 10.** Complexity of the Gröbner basis attacks for different number of variables $n$ and the strength of MQQ against Gröbner basis attacks.

### 5.7 Attack by an anonymous PKC-2008 reviewer

In this subsection we will describe an attack that has been proposed by an anonymous reviewer of the 11th International Workshop on Practice and Theory in Public Key Cryptography – PKC 2008, held in Barcelona, Spain in March 2008.

Let us take a simplified version of the algorithm where $P = T \circ P' \circ T$ (only one linear transformation $T$ is used) and in this subsection let us denote by $S$ the set of all $n$–bit Boolean vectors, i.e., $S \equiv \{0,1\}^n$. Thus, $P$, $P'$ and $T$ are now maps from $S$ to $S$, and elements of $S$ are the $n$–tuples $(x_1, x_2, ..., x_n)$, where $x_i \in \{0,1\}$.

The first step in the attack will be to identify an element $s$ of $S$, that under the linear transformation $T$ has the form:

$$T(s) = (0, 0, \dots, 0, x_{k+1}, x_{k+2}, \dots, x_n),$$

that is, such that its first $k$ coordinates are zero. Let $S_k$ be the space of elements of $S$ whose first $k$ entries are 0.

Selecting elements of $S$ at random, the probability of finding $s \in S_k$ is $2^{-k}$, so finding an element of $S_k$ reduces to being able to know that such an element has been found. To do this, notice that since the first $k$ coordinates of $T(s)$ are all zero, then **for every element** $r \in S$, the first $k$ coordinates of

$$P' \circ T(r)$$

and the first $k$ coordinates of

$$P' \circ T(r+s)$$

are equal, so the vector space generated by elements

$$\{P(r+s) - P(r) | r \in S\}$$

has co-dimension $k$ in S. This last condition can be easily verified.

The attack is as follows.

1. Pick a random $s \in S$.
2. **Check if $s$ is an element of $T^{-1}(S_k)$ using the previous test.**
3. Repeat until a basis for $T^{-1}(S_k)$ is found.

The reasons why we bolded the Step 2, the phrase "for every element" and several other phrases in the forthcoming text, will become clear at the end of this subsection.

If $T(s) \in S_k$, then we have seen that the first $k$ coordinates of $P' \circ T(s+r)$ coincide with the first $k$ coordinates of $P' \circ T(r)$ **for every** $r$, hence $P' \circ T(s+r) - P' \circ T(r)$ is an element of $S_k$ (**this is actually how one identifies that** $T(s) \in S_k$). Since $T$ is linear, $T(P' \circ T(s+r) - P' \circ T(r)) = P(s+r) - P(r)$. Notice that this is $T(s')$ for $s'$ an element of $S_k$. In other words, when one picks a random $s$ **and checks if $T(s)$ is an element of $S_k$** by checking that the co-dimension of $\{P(s+r) - P(r) \mid r \in S\}$ is $k$, this space of co-dimension $k$ is the same as $T(S_k)$.

Using a modification of the previous argument one can recursively identify $T^{-1}$ of the subspaces of $S$ such that first $k$ coordinates are zero. Again, as a by-product of this attack, the image of such subspaces under $T$ is recovered.

The attack described above is based on a very nice idea, namely to apply linear algebra theory in order to recover one part of the private key (the linear transformation $T$). However, the complexity of the attack is much worse than the projected security of our scheme that is $2^{\frac{n}{2}}$. The parts that we have bolded are clearly showing the complexity of the attack. Namely, in the first part of the attack (the Step 2) the attacker needs to find out an element $s$ such that it belongs to $T^{-1}(S_k)$. In order to do that, he pics a random $s$ and then he **have to check for every** $r \in S$ that $\{P(r+s) - P(r) \mid r \in S\}$ has co-dimension $k$ in S. The number of elements in $S$ is $2^n$. Thus, finding only one vector $s \in S_k$ needs $2^n$ operations, and recovering the whole basis of the discussed subspaces under $T$ will need many more operations. Moreover, the arguments that the attacker does not need to check all elements $r \in S$, but just a small number of random elements $r \in S$ that satisfy the condition that the first $k$ coordinates of $P(r+s)$ and $P(r)$ are equal is completely wrong and has no mathematical merit.

## 6 Conclusions

We have constructed a public key cryptosystem MQQ by using quasigroups. The main idea is to represent quasigroups as vector valued Boolean functions, to find a class of quasigroups that have degree at most 2, and then to use quasigroup string transformations to construct bijective trapdoor multivariate quadratic polynomials.

The speed of encryption/verification of our scheme is similar to other MQ schemes, and the speed of decryption/signing is in the range of 500–1000 times faster than the most popular public key schemes. Moreover, the algorithm offers flexibility in its implementation from parallelization point of view.

By learning about the weaknesses of all existing MQ schemas, we have designed a scheme that combines several known structures to create secure multivariate quadratic trapdoor functions.

Our scheme is resistant against known successful attacks on other MQ schemes, since its design principles does not include the known design weaknesses of other MQ schemes, one of which is that there exists a given multivariate quadratic polynomial (or a set of MQ polynomials) with relatively low degree, on top of which the whole PKC scheme is constructed. That design principle has been the necessary crib for successful attacks of many MQ schemes. In our PKC scheme, there is a fundamental difference in that we have defined a huge class of quasigroups called Multivariate Quadratic Quasigroups which gives MQ polynomials after certain types of operations.

It is an open research problem to count the number of MQQs of order $2^5$ and to find ways to construct MQQs of higher order (for example of order $2^8$). Finding such quasigroups will increase the security and will speed up the process of decryption/signing even more.

## References

1. M. Ajtai, "Generating hard instances of lattice problems", in *Proceedings of the 28th Annual ACM Symposium on Theory of Computing*, 1996, pp. 99–108.
2. G. Ars, J.-C. Faugère, H. Imai, M. Kawazoe and M . Sugita, "Comparison Between XL and Grobner basis Algorithms", in *Advances in Cryptology, ASIACRYPT 2004*, LNCS, Vol. 3329, pp. 338–353, 2004.
3. V. D. Belousov, *Osnovi teorii kvazigrup i lup* (in Russian), Nauka, Moscow, 1967.
4. A. Braeken, C. Wolf, and B. Preneel, "A study of the security of Unbalanced Oil and Vinegar signature schemes", in *The Cryptographer's Track at RSA Conference 2005*, LNCS, Vol. 3376, pp. 29–43, 2005. Available: `http://eprint.iacr.org/2004/222/`
5. J. F. Buss, G. S. Frandsen, and J. O. Shallit, "The computational complexity of some problems of linear algebra", Research Series RS-96-33, BRICS, Department of Computer Science, University of Aarhus, September 1996. Available: `http://www.brics.dk/RS/96/33/`.
6. D. Coppersmith, J. Stern, and S. Vaudenay, "Attacks on the birational permutation signature schemes", in *Advances in Cryptology, CRYPTO 1993*, LNCS, Vol. 773, pp. 435–443, 1993.

7. N. Courtois, A. Klimov, J. Patarin, and A. Shamir, "Efficient Algorithms for Solving Overdefined Systems of Multivariate Polynomial Equations", in *Advances in Cryptology, EUROCRYPT 2000*, LNCS, Vol. 1807, pp. 392–407, 2000.
8. J. Dénes, A. D. Keedwell, *Latin Squares and their Applications*, English Univer. Press Ltd., 1974.
9. W. Diffie and M. Hellman, "New Directions in Cryptography", *IEEE Trans. Information Theory*, Vol. IT-22, No 6, (1976), 644–654.
10. J. Ding, "A New Variant of the Matsumoto-Imai Cryptosystem through Perturbation", in *PKC04*, LNCS Vol. 2947, pp. 305-318, 2004.
11. H. Dobbertin, "One-to-one highly nonlinear power functions on $GF(2^n)$", *Appl. Algebra Eng. Commun. Comput.*, Vol. 9(2), pp. 139-152, 1998.
12. V. Dubois, P.A. Fouque, A. Shamir and J. Stern, "Practical Cryptanalysis of SFLASH", in *Advances in Cryptology CRYPTO 2007*, LNCS, Springer-Verlag, 2007, to be published.
13. "D.VAM.9-1.1: Report on eBATS Performance Benchmarks." Comprehensive report on measurements collected in the first stage of eBATS, March 2007. Available: `http://www.ecrypt.eu.org/ebats/`
14. J.-C. Faugère, "A new efficient algorithm for computing Gröbner basis ($F4$)". Available: `http://citeseer.ist.psu.edu/faugere00new.html`
15. J.-C. Faugère, A. Joux, "Algebraic Cryptanalysis of Hidden Field Equation (HFE) Cryptosystems Using Gröbner Bases", in *Advances in Cryptology CRYPTO 2003*, LNCS, Vol. 2729, pp 44–60, 2003.
16. P.-A. Fouque, L. Granboulan and J. Stern, "Differential Cryptanalysis for Multivariate Schemes", in *Advances in Cryptology EUROCRYPT 2005*, LNCS Vol. 3494, pp. 341–353, 2005.
17. W. Geiselmann, W. Meier, and R. Steinwandt, "An Attack on the Isomorphisms of Polynomials Problem with One Secret", *Int. Journal of Information Security*, Vol. 2(1), pp. 59-64, 2003.
18. O. Goldreich, S. Goldwasser and S. Halevi, "Public-Key Cryptosystems from Lattice Reduction Problems", LNCS, Vol. 1294, pp. 112–131, 1997.
19. L. Goubin and N. T. Courtois, "Cryptanalysis of the TTM cryptosystem", in *Advances in Cryptology, ASIACRYPT 2000*, LNCS Vol. 1976, pp. 44–57, 2000.
20. Helion Technology Limited, "Modular Exponentiation Engine for RSA and DH (ModExp)", February 16, 2007, Available: `http://www.xilinx.com/publications/3rd_party/products/Helion_ModExp_AllianceCORE_data_sheet.pdf`
21. J. Hoffstein, J. Pipher and J. H. Silverman, "NTRU: A ring based public key cryptosystem", LNCS, Vol. 1433 pp. 267–288, 1998.
22. H. Imai and T. Matsumoto, "Algebraic methods for constructing asymmetric cryptosysytems", in *Proceedings of 3rd Intern. Conf. AAECC-3*, LNCS Vol. 29, pp. 108–119, 1985.
23. A. Kipnis, J. Patarin, and L. Goubin, "Unbalanced Oil and Vinegar signature schemes", in *Advances in Cryptology, EUROCRYPT 1999*, LNCS Vol. 1592, pp. 206–222, 1999.
24. A. Kipnis, A. Shamir, "Cryptanalysis of the HFE public key cryptosystem", in *Advances in Cryptology, CRYPTO 1999*, LNCS Vol. 1666, pp. 19–30, 1999.
25. K.H. Ko, S.J. Lee, J.H. Cheon, J.W. Han, J. Kang C. Park, "New Public-Key Cryptosystem Using Braid Groups", in *Advances in Cryptology, CRYPTO 2001*, LNCS, Vol. 1880, pp. 166–184, 2001.
26. N. Koblitz, "Elliptic curve cryptosystems", in *Mathematics of Computation 48*, 1987, pp. 203–209.
27. F. Levy-dit-Vehel, and L. Perret, "Polynomial equivalence problems and applications to multivariate cryptosystems", in *Progress in Cryptology - INDOCRYPT 2003*, LNCS Vol. 2904, pp. 235-251, 2003.
28. S. Markovski, D. Gligoroski, V. Bakeva, "Quasigroup String Processing: Part 1", Maced. Acad. of Sci. and Arts, Sc. Math. Tech. Scien. XX 1-2, pp. 13–28, 1999.
29. S. Markovski, "Quasigroup string processing and applications in cryptography", in *Proc. 1-st Inter. Conf. Mathematics and Informatics for industry – MII, Thessaloniki 2003*, pp. 278–290, 2003.
30. R.J. McEliece, "A Public-key cryptosystem based on algebraic coding theory", DSN Progress Report, Jet Propulsion Laboratory, Pasadena, CA, 1978, pp. 114–116.
31. V. Miller, "Use of elliptic curves in cryptography", in *CRYPTO 85*, 1985.
32. T. Moh, "A public key system with signature and master key function", *Communications in Algebra*, 27(5), pp. 2207–2222, 1999. Available: `http://citeseer/moh99public.html`
33. J. Patarin, "Cryptanalysis of the Matsumoto and Imai public key scheme of Eurocrypt'88", in *Advances in Cryptology, CRYPTO 1995*, LNCS, Vol. 963, pp. 248–261, 1995.
34. J. Patarin, "Hidden Field Equations (HFE) and Isomorphisms of Polynomials (IP): two new families of asymmetric algorithms", in *Advances in Cryptology, EUROCRYPT 1996*, LNCS Vol. 1070, pp. 33–48, 1996.
35. J. Patarin, "Hidden Field Equations (HFE) and Isomorphisms of Polynomials (IP): two new families of asymmetric algorithms - EXTENDED VERSION OF THE EUROCRYPT 1996 paper". Available from the author.

36. J. Patarin, "The oil and vinegar signature scheme", presented at the Dagstuhl Workshop on Cryptography, September 1997.
37. J. Patarin, N. Courtois, and L. Goubin, "FLASH, a Fast Multivariate Signature Algorithm", in *CT-RSA 01*, LNCS Vol. 2020, pp. 297-307, 2001.
38. L. Peret, "A Fast Cryptanalysis of the Isomorphism of Polynomials with One Secret Problem", in *EUROCRYPT 2005*, LNCS Vol. 3494, pp. 354-370, 2005.
39. M.O. Rabin, "Digital Signatures and Public-Key Functions as Intractable as Factorization", Technical Report MIT/LCS/TR-212, M.I.T., 1978.
40. R. Rivest, A. Shamir and L. Adleman, "A Method for Obtaining Digital Signatures and Public Key Cryptosystems", *Comm. ACM*, Vol. 21, No 2, 1978, pp. 120–126.
41. A. Shamir, "Efficient signature schemes based on birational permutations", in *Advances in Cryptology, CRYPTO 1993*, LNCS, Vol. 773, pp. 1–12, 1993.
42. J. D. H. Smith, *An introduction to quasigroups and their representations*, Chapman & Hall/CRC, ISBN 1-58488-537-8, 2007.
43. M. Sugita, M. Kawazoe and H. Imai, "Relation between the XL Algorithm and Grobner Basis Algorithms", IEICE-Tran Fund Elec, Comm & Comp Sci Vol. E89-A, Number 1 pp. 11–18, 2006.
44. C. Wolf, A. Braeken, and B. Preneel, "Efficient cryptanalysis of RSE(2)PKC and RSSE(2)PKC", in *Conference on Security in Communication Networks – SCN 2004*, LNCS, Vol. 3352, pp. 294-309, 2005. Available extended version: `http://eprint.iacr.org/2004/237`.
45. C. Wolf and B. Preneel, "Asymmetric cryptography: Hidden Field Equations", in *European Congress on Computational Methods in Applied Sciences and Engineering 2004*, Jyväskylä University, 2004. Available extended version: `http://eprint.iacr.org/2004/072/`
46. C. Wolf and B. Preneel, "Equivalent keys in HFE, $C^*$, and variations", Cryptology ePrint Archive, Report 2004/360, 2004. Available: `http://eprint.iacr.org/2004/360/`
47. C. Wolf and B. Preneel, "Taxonomy of Public Key Schemes based on the problem of Multivariate Quadratic equations", Cryptology ePrint Archive, Report 2005/077, 2005. Available: `http://eprint.iacr.org/`
48. C. Wolf and B. Preneel, "Superfluous keys in Multivariate Quadratic asymmetric systems", in *Public Key Cryptography–PKC 2005*, LNCS Vol. 3386, pp. 275–287, 2005.

# APPENDIX

## AN EXAMPLE OF THE CREATION OF A PRIVATE AND A PUBLIC KEY WITH $n = 20$ BITS

This example is for $n = 20$. Since even with such a small example, the number of terms in some expressions will increase to more than 100, in the notation we will use horizontal lines to make a distinction between different coordinates.

We will use the simplified version of the algorithm where $P = T \circ P' \circ T$. Let $x = (x_1, x_2, \ldots, x_{20})$ be a vector of 20 Boolean variables. The private and the public key is created by the following procedure:

1) Set $T =$

$$\begin{bmatrix}
0&0&0&1&1&0&1&0&0&0&1&0&0&1&1&1&0&0&0&1\\
1&0&0&1&1&1&1&1&0&1&1&1&1&1&1&0&0&0&1&1\\
1&1&1&1&0&1&1&1&1&1&0&0&1&1&1&1&1&1&1&0\\
0&0&1&1&1&0&0&0&1&1&1&1&0&1&0&0&1&0&1&0\\
1&1&1&0&0&0&1&1&0&1&0&0&0&0&1&0&1&0&1&1\\
1&1&1&1&1&1&1&1&1&1&1&0&1&1&0&1&1&1&1&1\\
0&1&0&1&1&1&0&0&1&1&1&1&0&0&1&0&0&0&0&0\\
0&0&0&0&0&1&1&1&0&1&0&1&1&1&0&0&0&1&0&0\\
0&1&0&0&0&0&1&1&1&1&1&1&0&1&1&1&1&0&0&1\\
1&1&1&1&1&1&1&0&1&0&0&0&1&0&0&0&1&1&0&1\\
0&1&0&1&0&1&1&1&1&1&0&0&0&0&0&0&0&1&0&1\\
0&0&1&1&0&1&0&1&0&1&0&0&1&0&0&1&1&0&1&0\\
0&1&1&1&0&0&0&0&0&1&0&0&0&1&0&1&0&1&0&0\\
1&0&1&0&0&0&1&0&0&1&1&1&1&0&1&1&1&0&1&1\\
0&0&1&1&1&1&0&0&0&1&1&1&1&1&1&0&0&1&0&0\\
0&1&1&1&1&0&0&0&0&1&0&0&0&1&0&0&0&1&1&1\\
0&0&0&0&0&1&0&1&0&1&0&1&0&1&1&1&0&0&0&1\\
0&0&0&1&1&0&1&0&0&1&0&1&0&1&1&0&1&1&1&1\\
1&1&0&0&0&1&0&1&0&0&0&1&1&0&1&0&1&1&1&1\\
0&1&0&1&1&1&1&1&1&0&1&0&0&0&1&0&1&1&0&1
\end{bmatrix},$$

where $T$ is a nonsingular $20 \times 20$ Boolean matrix generated uniformly at random;

2) Set

$*_1(x_1, x_2, x_3, x_4, x_5, x_6, x_7, x_8, x_9, x_{10}) =$

$$\left[\begin{array}{l}
1 + x_1 + x_2 + x_3 + x_7 + x_9 + x_{10} \\ \hline
1 + x_6x_1 + x_8x_1 + x_{10}x_1 + x_1 + x_2 + x_4 + x_2x_7 + x_3x_7 + x_4x_7 + \\
+ x_5x_7 + x_2x_8 + x_3x_8 + x_4x_8 + x_2x_9 + x_3x_9 + x_5x_9 + x_9 + x_5x_{10} + \\
+ x_{10} \\ \hline
1 + x_6x_1 + x_9x_1 + x_1 + x_2x_6 + x_3x_6 + x_6 + x_4x_7 + x_7 + x_4x_8 + \\
+ x_8 + x_2x_9 + x_3x_9 + x_4x_{10} \\ \hline
1 + x_8x_4 + x_9x_4 + x_4 + x_5 + x_1x_6 + x_2x_6 + x_3x_6 + x_1x_7 + x_2x_7 + \\
+ x_3x_7 + x_1x_8 + x_2x_8 + x_3x_8 + x_8 + x_9 + x_1x_{10} + x_2x_{10} + x_3x_{10} + \\
+ x_{10} \\ \hline
1 + x_7x_1 + x_{10}x_1 + x_1 + x_2 + x_3 + x_4 + x_2x_7 + x_3x_7 + x_8 + x_9 + \\
+ x_2x_{10} + x_3x_{10}
\end{array}\right]^T,$$

$*_2(x_1, x_2, x_3, x_4, x_5, x_6, x_7, x_8, x_9, x_{10}) =$

$$\left[\begin{array}{l}
x_1 + x_2 + x_6 + x_9 + x_{10} \\ \hline
x_8x_2 + x_2 + x_3 + x_3x_6 + x_5x_6 + x_6 + x_3x_7 + x_4x_7 + x_5x_7 + \\
+ x_1x_8 + x_3x_8 + x_4x_8 + x_5x_8 + x_8 + x_3x_9 + x_5x_9 + x_3x_{10} + \\
+ x_5x_{10} + x_{10} \\ \hline
1 + x_9x_1 + x_{10}x_1 + x_1 + x_2 + x_4 + x_5 + x_3x_7 + x_4x_7 + x_8 + x_2x_9 + \\
+ x_2x_{10} \\ \hline
x_7x_3 + x_9x_3 + x_{10}x_3 + x_3 + x_4 + x_5 + x_1x_6 + x_2x_6 + x_4x_7 + x_7 + \\
+ x_1x_8 + x_2x_8 + x_4x_9 + x_4x_{10} \\ \hline
x_3 + x_4 + x_1x_6 + x_2x_6 + x_7 + x_8 + x_1x_9 + x_2x_9 + x_{10}
\end{array}\right]^T,$$

$*_3(x_1, x_2, x_3, x_4, x_5, x_6, x_7, x_8, x_9, x_{10}) =$

$$\left[\begin{array}{l}
x_7x_1 + x_9x_1 + x_1 + x_2 + x_4 + x_3x_6 + x_4x_6 + x_5x_6 + x_2x_7+ \\
+x_3x_7 + x_4x_7 + x_5x_7 + x_2x_8 + x_3x_8 + x_4x_8 + x_5x_8 + x_3x_9+ \\
+x_4x_9 + x_5x_9 + x_9 + x_2x_{10} + x_{10} \\ \hline
x_6x_1 + x_7x_1 + x_9x_1 + x_1 + x_3 + x_5 + x_2x_6 + x_6 + x_2x_7+ \\
+x_2x_8 + x_2x_{10} + x_3x_{10} + x_4x_{10} + x_5x_{10} + x_{10} \\ \hline
1 + x_6x_1 + x_7x_1 + x_9x_1 + x_{10}x_1 + x_1 + x_4 + x_2x_6 + x_2x_7+ \\
+x_7 + x_2x_8 + x_8 + x_9 + x_3x_{10} + x_4x_{10} + x_5x_{10} + x_{10} \\ \hline
1 + x_9x_1 + x_{10}x_1 + x_1 + x_4 + x_5 + x_2x_6 + x_6 + x_3x_7 + x_4x_7+ \\
+x_5x_7 + x_2x_{10} + x_3x_{10} + x_4x_{10} + x_5x_{10} \\ \hline
1 + x_6x_2 + x_{10}x_2 + x_2 + x_4 + x_5 + x_3x_7 + x_4x_7 + x_5x_7+ \\
+x_7 + x_1x_9 + x_9 + x_1x_{10} + x_3x_{10} + x_4x_{10} + x_5x_{10} + x_{10}
\end{array}\right]^T ;$$

3) The tuple $(T, *_1, *_2, *_3)$ is the private key;

4) Set $x' = T \cdot x^T =$

$$\left[\begin{array}{l}
x_4 + x_5 + x_7 + x_{11} + x_{14} + x_{15} + x_{16} + x_{20} \\ \hline
x_1 + x_4 + x_5 + x_6 + x_7 + x_8 + x_{10} + x_{11} + x_{12}+ \\
+x_{13} + x_{14} + x_{15} + x_{19} + x_{20} \\ \hline
x_1 + x_2 + x_3 + x_4 + x_6 + x_7 + x_8 + x_9 + x_{10}+ \\
+x_{13} + x_{14} + x_{15} + x_{16} + x_{17} + x_{18} + x_{19} \\ \hline
x_3 + x_4 + x_5 + x_9 + x_{10} + x_{11} + x_{12} + x_{14} + x_{17} + x_{19} \\ \hline
x_1 + x_2 + x_3 + x_7 + x_8 + x_{10} + x_{15} + x_{17} + x_{19} + x_{20} \\ \hline
x_1 + x_2 + x_3 + x_4 + x_5 + x_6 + x_7 + x_8 + x_9 + x_{10} + x_{11}+ \\
+x_{13} + x_{14} + x_{16} + x_{17} + x_{18} + x_{19} + x_{20} \\ \hline
x_2 + x_4 + x_5 + x_6 + x_9 + x_{10} + x_{11} + x_{12} + x_{15} \\ \hline
x_6 + x_7 + x_8 + x_{10} + x_{12} + x_{13} + x_{14} + x_{18} \\ \hline
x_2 + x_7 + x_8 + x_9 + x_{10} + x_{11} + x_{12} + x_{14} + x_{15}+ \\
+x_{16} + x_{17} + x_{20} \\ \hline
x_1 + x_2 + x_3 + x_4 + x_5 + x_6 + x_7 + x_9 + x_{13}+ \\
+x_{17} + x_{18} + x_{20} \\ \hline
x_2 + x_4 + x_6 + x_7 + x_8 + x_9 + x_{10} + x_{18} + x_{20} \\ \hline
x_3 + x_4 + x_6 + x_8 + x_{10} + x_{13} + x_{16} + x_{17} + x_{19} \\ \hline
x_2 + x_3 + x_4 + x_{10} + x_{14} + x_{16} + x_{18} \\ \hline
x_1 + x_3 + x_7 + x_{10} + x_{11} + x_{12} + x_{13} + x_{15} + x_{16}+ \\
+x_{17} + x_{19} + x_{20} \\ \hline
x_3 + x_4 + x_5 + x_6 + x_{10} + x_{11} + x_{12} + x_{13} + x_{14}+ \\
+x_{15} + x_{18} \\ \hline
x_2 + x_3 + x_4 + x_5 + x_{10} + x_{14} + x_{18} + x_{19} + x_{20} \\ \hline
x_6 + x_8 + x_{10} + x_{12} + x_{14} + x_{15} + x_{16} + x_{20} \\ \hline
x_4 + x_5 + x_7 + x_{10} + x_{12} + x_{14} + x_{15} + x_{17} + x_{18}+ \\
+x_{19} + x_{20} \\ \hline
x_1 + x_2 + x_6 + x_8 + x_{12} + x_{13} + x_{15} + x_{17} + x_{18}+ \\
+x_{19} + x_{20} \\ \hline
x_2 + x_4 + x_5 + x_6 + x_7 + x_8 + x_9 + x_{11} + x_{15}+ \\
+x_{17} + x_{18} + x_{20}
\end{array}\right]^T ;$$

5) Represent the vector $x'$ by chunks of 5 bits, i.e. $x' = X_1X_2X_3X_4$;

6) Compute $y' = Y_1Y_2Y_3Y_4$ such that $Y_1 = X_1$, $Y_2 = X_1 *_1 X_2$, $Y_3 = X_2 *_2 X_3$ and $Y_4 = X_3 *_3 X_4$. The following relations will be obtained:

$Y_1 = (y_{11}, y_{12}, y_{13}, y_{14}, y_{15})$, where

$$y_{11} = x_4 + x_5 + x_7 + x_{11} + x_{14} + x_{15} + x_{16} + x_{20},$$
$$y_{12} = x_1 + x_4 + x_5 + x_6 + x_7 + x_8 + x_{10} + x_{11} + x_{12} + +x_{13} + x_{14} + x_{15} + x_{19} + x_{20},$$
$$y_{13} = x_1 + x_2 + x_3 + x_4 + x_6 + x_7 + x_8 + x_9 + x_{10} + +x_{13} + x_{14} + x_{15} + x_{16} + x_{17} + x_{18} + x_{19},$$
$$y_{14} = x_3 + x_4 + x_5 + x_9 + x_{10} + x_{11} + x_{12} + x_{14} + x_{17} + x_{19},$$
$$y_{15} = x_1 + x_2 + x_3 + x_7 + x_8 + x_{10} + x_{15} + x_{17} + x_{19} + x_{20},$$

$Y_2 = (y_{21}, y_{22}, y_{23}, y_{24}, y_{25})$, where

$$y_{21} = 1 + x_1 + x_4 + x_7 + x_8 + x_{12} + x_{13} + x_{15} + x_{16} + x_{17},$$
$$y_{22} = 1 + x_7x_1 + x_9x_1 + x_{10}x_1 + x_{13}x_1 + x_{14}x_1 + x_{15}x_1 + x_{16}x_1 + x_{17}x_1 + x_{18}x_1 + x_{19}x_1 + x_{20}x_1 + x_1 + x_2 + x_2x_3 + x_3 + x_4 + x_4x_5 + x_4x_6 + x_5x_6 + x_2x_7 + x_6x_7 + x_7 + x_3x_8 + x_4x_8 + x_5x_8 + x_7x_8 + x_8 + x_4x_9 + x_5x_9 + x_2x_{10} + x_4x_{10} + x_7x_{10} + x_9x_{10} + x_{10} + x_2x_{11} + x_3x_{11} + x_6x_{11} + x_7x_{11} + x_8x_{11} + x_3x_{12} + x_4x_{12} + x_8x_{12} + x_{10}x_{12} + x_{11}x_{12} + x_3x_{13} + x_5x_{13} + x_8x_{13} + x_{11}x_{13} + x_6x_{14} + x_7x_{14} + x_{11}x_{14} + x_{14} + x_2x_{15} + x_4x_{15} + x_5x_{15} + x_6x_{15} + x_7x_{15} + x_8x_{15} + x_{10}x_{15} + x_{15} + x_5x_{16} + x_6x_{16} + x_7x_{16} + x_8x_{16} + x_9x_{16} + x_{11}x_{16} + x_{12}x_{16} + x_{14}x_{16} + x_2x_{17} + x_5x_{17} + x_7x_{17} + x_9x_{17} + x_{10}x_{17} + x_{13}x_{17} + x_{15}x_{17} + x_{16}x_{17} + x_3x_{18} + +x_4x_{18} + x_8x_{18} + x_{10}x_{18} + x_{11}x_{18} + x_{12}x_{18} + x_{13}x_{18} + x_{16}x_{18} + x_3x_{19} + x_{14}x_{19} + x_4x_{20} + x_6x_{20} + x_7x_{20} + x_8x_{20} + x_{10}x_{20} + x_{12}x_{20} + x_{13}x_{20} + x_{14}x_{20} + x_{19}x_{20} + x_{20},$$
$$y_{23} = 1 + x_2x_1 + x_5x_1 + x_7x_1 + x_{10}x_1 + x_{11}x_1 + x_{15}x_1 + x_{18}x_1 + x_{19}x_1 + x_1 + x_2x_3 + x_3 + x_2x_4 + x_3x_4 + x_2x_5 + x_4x_5 + x_5 + x_2x_6 + x_5x_6 + x_6 + x_3x_7 + x_4x_7 + x_5x_7 + x_6x_7 + x_7 + x_3x_8 + x_4x_8 + x_5x_8 + x_4x_9 + x_5x_9 + x_8x_9 + x_3x_{10} + x_6x_{10} + x_8x_{10} + x_{10} + x_4x_{11} + x_5x_{11} + x_6x_{11} + x_8x_{11} + x_9x_{11} + x_{10}x_{11} + x_2x_{12} + x_3x_{12} + x_5x_{12} + x_7x_{12} + x_8x_{12} + x_9x_{12} + x_{11}x_{12} + x_{12} + x_2x_{13} + x_3x_{13} + x_4x_{13} + x_7x_{13} + x_9x_{13} + x_{12}x_{13} + x_3x_{14} + x_8x_{14} + x_9x_{14} + x_{10}x_{14} + x_{13}x_{14} + x_2x_{15} + x_3x_{15} + x_4x_{15} + x_6x_{15} + x_7x_{15} + x_{10}x_{15} + x_{11}x_{15} + x_{12}x_{15} + x_{13}x_{15} + x_{15} + x_3x_{17} + x_8x_{17} + x_9x_{17} + x_{10}x_{17} + x_{13}x_{17} + x_2x_{18} + x_5x_{18} + x_6x_{18} + x_7x_{18} + x_9x_{18} + x_{13}x_{18} + x_{14}x_{18} + x_{17}x_{18} + x_{18} + x_2x_{19} + x_4x_{19} + x_6x_{19} + x_7x_{19} + x_8x_{19} + x_9x_{19} + x_{11}x_{19} + x_{12}x_{19} + x_{18}x_{19} + x_{19} + x_3x_{20} + x_4x_{20} + x_5x_{20} + x_9x_{20} + x_{10}x_{20} + x_{11}x_{20} + x_{12}x_{20} + x_{14}x_{20} + x_{17}x_{20} + x_{19}x_{20},$$
$$y_{24} = 1+x_4x_3+x_5x_3+x_6x_3+x_7x_3+x_{10}x_3+x_{11}x_3+x_{17}x_3+x_{19}x_3+x_{20}x_3+x_3+x_2x_4+x_4+x_4x_5+x_4x_6+x_5x_6+x_5x_7+x_7+x_8+x_2x_9+x_4x_9+x_6x_9+x_9+x_4x_{10}+x_6x_{10}+x_7x_{10}+x_2x_{11}+x_4x_{11}+x_5x_{11}+x_6x_{11}+x_{10}x_{11}+x_{11}+x_4x_{12}+x_5x_{12}+x_6x_{12}+x_7x_{12}+x_{10}x_{12}+x_{11}x_{12}+x_{12}+x_2x_{13}+x_5x_{13}+x_7x_{13}+x_{10}x_{13}+x_{11}x_{13}+x_4x_{14}+x_5x_{14}+x_6x_{14}+x_7x_{14}+x_{10}x_{14}+x_{11}x_{14}+x_{14}+x_4x_{15}+x_9x_{15}+x_{11}x_{15}+x_{13}x_{15}+x_{15}+x_2x_{16}+x_5x_{16}+x_7x_{16}+x_{10}x_{16}+x_{11}x_{16}+x_{15}x_{16}+x_{16}+x_6x_{17}+x_7x_{17}+x_9x_{17}+x_{12}x_{17}+x_{14}x_{17}+x_{17}+x_4x_{18}+x_9x_{18}+x_{11}x_{18}+x_{13}x_{18}+x_{16}x_{18}+x_{18}+x_4x_{19}+x_6x_{19}+x_7x_{19}+x_{11}x_{19}+x_{12}x_{19}+x_{13}x_{19}+x_{14}x_{19}+x_{16}x_{19}+x_4x_{20}+x_5x_{20}+x_9x_{20}+x_{10}x_{20}+x_{11}x_{20}+x_{12}x_{20}+x_{14}x_{20}+x_{17}x_{20}+x_{19}x_{20}+x_{20},$$
$$y_{25} = 1 + x_1x_2 + x_3x_2 + x_7x_2 + x_{10}x_2 + x_{11}x_2 + x_{12}x_2 + x_{13}x_2 + x_{15}x_2 + x_{17}x_2 + x_{18}x_2 + x_{20}x_2 + x_1x_3 + x_3 + x_1x_4 + x_3x_4 + x_5 + x_6 + x_1x_7 + x_4x_7 + x_1x_9 + x_3x_9 + x_7x_9 + x_9 + x_3x_{10} + x_4x_{10} + x_7x_{10} + x_9x_{10} + x_{10} + x_3x_{11} + x_4x_{11} + x_7x_{11} + x_9x_{11} + x_1x_{12} + x_4x_{12} + x_9x_{12} + x_{10}x_{12} + x_{11}x_{12} + x_{12} + x_3x_{13} + x_4x_{13} + x_7x_{13} + x_9x_{13} + x_{12}x_{13} + x_{13} + x_1x_{14} + x_3x_{14} + x_7x_{14} + x_{10}x_{14} + x_{11}x_{14} + x_{12}x_{14} + x_{13}x_{14} + x_1x_{15} + x_4x_{15} + x_9x_{15} + x_{10}x_{15} + x_{11}x_{15} + x_{13}x_{15} + x_{14}x_{15} + x_{15} + x_{16} + x_1x_{17} + x_4x_{17} + x_9x_{17} + x_{10}x_{17} + x_{11}x_{17} + x_{13}x_{17} + x_{14}x_{17} + x_1x_{18} + x_4x_{18} + x_9x_{18} + x_{10}x_{18} + x_{11}x_{18} + x_{13}x_{18} + x_{14}x_{18} + x_{18} + x_{19} + x_3x_{20} + x_4x_{20} + x_7x_{20} + x_9x_{20} + x_{12}x_{20} + x_{14}x_{20} + x_{15}x_{20} + x_{17}x_{20} + x_{18}x_{20} + x_{20},$$

$Y_3 = (y_{31}, y_{32}, y_{33}, y_{34}, y_{35})$, where

$$y_{31} = x_2 + x_3 + x_5 + x_7 + x_9 + x_{10} + x_{12} + x_{13} + x_{15} + x_{18} + x_{20},$$
$$y_{32} = x_6x_1+x_8x_1+x_{10}x_1+x_{12}x_1+x_{13}x_1+x_{17}x_1+x_{20}x_1+x_1+x_2x_3+x_3+x_3x_4+x_2x_5+x_3x_5+x_4x_5+x_5+x_3x_6+x_4x_6+x_5x_6+x_3x_7+x_4x_7+x_6x_7+x_2x_8+x_3x_8+x_4x_8+x_5x_8+x_7x_8+x_8+x_3x_9+x_4x_9+x_9+x_4x_{10}+x_7x_{10}+x_8x_{10}+x_2x_{11}+x_6x_{11}+x_8x_{11}+x_3x_{12}+x_4x_{12}+x_5x_{12}+x_8x_{12}+x_9x_{12}+x_{12}+x_2x_{13}+x_5x_{13}+x_7x_{13}+x_{10}x_{13}+x_{11}x_{13}+x_2x_{14}+x_4x_{14}+x_5x_{14}+x_8x_{14}+x_{10}x_{14}+x_{11}x_{14}+x_{13}x_{14}+x_{14}+x_3x_{15}+x_4x_{15}+x_6x_{15}+x_8x_{15}+x_{10}x_{15}+x_{13}x_{15}+x_2x_{16}+x_3x_{16}+x_5x_{16}+x_6x_{16}+x_7x_{16}+x_8x_{16}+x_9x_{16}+x_{15}x_{16}+x_2x_{17}+x_3x_{17}+x_4x_{17}+x_5x_{17}+x_7x_{17}+x_{10}x_{17}+x_{11}x_{17}+x_{12}x_{17}+x_{14}x_{17}+x_{15}x_{17}+x_{16}x_{17}+x_{17}+x_2x_{18}+x_3x_{18}+x_5x_{18}+x_8x_{18}+x_{10}x_{18}+x_{11}x_{18}+x_{12}x_{18}+x_{13}x_{18}+x_{16}x_{18}+x_{17}x_{18}+x_{18}+x_3x_{19}+x_4x_{19}+x_7x_{19}+x_8x_{19}+x_9x_{19}+x_{11}x_{19}+x_{12}x_{19}+x_{15}x_{19}+x_{17}x_{19}+x_{18}x_{19}+x_3x_{20}+x_4x_{20}+x_5x_{20}+x_8x_{20}+x_9x_{20}+x_{17}+x_{20}+x_{18}x_{20}+x_{19}x_{20}+x_{20},$$

$$y_{33} = 1+x_3x_1+x_4x_1+x_5x_1+x_6x_1+x_8x_1+x_{12}x_1+x_{13}x_1+x_{15}x_1+x_1+x_2+x_2x_3+x_3+x_2x_4+x_3x_4+x_3x_5+x_5+x_2x_6+x_4x_6+x_3x_7+x_4x_7+x_5x_7+x_6x_7+x_2x_8+x_4x_8+x_5x_8+x_7x_8+x_3x_9+x_4x_9+x_6x_9+x_8x_9+x_2x_{10}+x_6x_{10}+x_9x_{10}+x_3x_{11}+x_4x_{11}+x_6x_{11}+x_8x_{11}+x_{10}x_{11}+x_{11}+x_4x_{12}+x_5x_{12}+x_6x_{12}+x_7x_{12}+x_2x_{13}+x_3x_{13}+x_5x_{13}+x_6x_{13}+x_7x_{13}+x_8x_{13}+x_9x_{13}+x_{10}x_{13}+x_{11}x_{13}+x_{13}+x_3x_{14}+x_4x_{14}+x_5x_{14}+x_6x_{14}+x_8x_{14}+x_{12}x_{14}+x_{13}x_{14}+x_3x_{15}+x_5x_{15}+x_7x_{15}+x_8x_{15}+x_{10}x_{15}+x_{13}x_{15}+x_{14}x_{15}+x_2x_{16}+x_5x_{16}+x_6x_{16}+x_9x_{16}+x_{10}x_{16}+x_{11}x_{16}+x_{12}x_{16}+x_{13}x_{16}+x_{16}+x_2x_{17}+x_5x_{17}+x_6x_{17}+x_9x_{17}+x_{10}x_{17}+x_{11}x_{17}+x_{12}x_{17}+x_{13}x_{17}+x_{17}+x_5x_{18}+x_{10}x_{18}+x_{12}x_{18}+x_{15}x_{18}+x_{16}x_{18}+x_{17}x_{18}+x_2x_{19}+x_3x_{19}+x_4x_{19}+x_5x_{19}+x_8x_{19}+x_9x_{19}+x_{11}x_{19}+x_{12}x_{19}+x_{16}x_{19}+x_{17}x_{19}+x_{18}x_{19}+x_5x_{20}+x_{10}x_{20}+x_{12}x_{20}+x_{15}x_{20}+x_{16}x_{20}+x_{17}x_{20}+x_{19}x_{20},$$

$$y_{34} = x_2x_1+x_3x_1+x_7x_1+x_8x_1+x_{11}x_1+x_{13}x_1+x_{14}x_1+x_{15}x_1+x_{17}x_1+x_{18}x_1+x_1+x_2x_3+x_3+x_2x_5+x_5+x_5x_6+x_6+x_2x_7+x_2x_8+x_5x_9+x_2x_{10}+x_6x_{10}+x_9x_{10}+x_{10}+x_3x_{11}+x_5x_{11}+x_7x_{11}+x_8x_{11}+x_{10}x_{11}+x_{11}+x_3x_{12}+x_6x_{12}+x_7x_{12}+x_8x_{12}+x_9x_{12}+x_2x_{13}+x_5x_{13}+x_{10}x_{13}+x_{11}x_{13}+x_2x_{14}+x_{11}x_{14}+x_{12}x_{14}+x_{14}+x_5x_{15}+x_6x_{15}+x_9x_{15}+x_{10}x_{15}+x_{13}x_{15}+x_{15}+x_3x_{16}+x_5x_{16}+x_6x_{16}+x_7x_{16}+x_8x_{16}+x_9x_{16}+x_{10}x_{16}+x_{12}x_{16}+x_{14}x_{16}+x_{15}x_{16}+x_{16}+x_5x_{17}+x_6x_{17}+x_9x_{17}+x_{10}x_{17}+x_{13}x_{17}+x_{16}x_{17}+x_{17}+x_2x_{18}+x_5x_{18}+x_{10}x_{18}+x_{11}x_{18}+x_{15}x_{18}+x_{17}x_{18}+x_{18}+x_3x_{19}+x_6x_{19}+x_7x_{19}+x_8x_{19}+x_9x_{19}+x_{14}x_{19}+x_{16}x_{19}+x_{19}+x_2x_{20}+x_3x_{20}+x_5x_{20}+x_7x_{20}+x_8x_{20}+x_{10}x_{20}+x_{11}x_{20}+x_{12}x_{20}+x_{13}x_{20}+x_{14}x_{20}+x_{16}x_{20}+x_{18}x_{20}+x_{19}x_{20},$$

$$y_{35} = x_2x_1+x_4x_1+x_6x_1+x_7x_1+x_9x_1+x_{11}x_1+x_{14}x_1+x_{20}x_1+x_1+x_2x_3+x_3x_4+x_4+x_5+x_3x_6+x_6+x_2x_7+x_3x_7+x_4x_7+x_6x_7+x_2x_8+x_4x_8+x_6x_8+x_7x_8+x_3x_9+x_7x_9+x_8x_9+x_9+x_{10}+x_3x_{11}+x_7x_{11}+x_8x_{11}+x_2x_{12}+x_4x_{12}+x_6x_{12}+x_7x_{12}+x_9x_{12}+x_{11}x_{12}+x_2x_{13}+x_4x_{13}+x_6x_{13}+x_7x_{13}+x_9x_{13}+x_{11}x_{13}+x_2x_{14}+x_3x_{14}+x_4x_{14}+x_6x_{14}+x_8x_{14}+x_9x_{14}+x_{11}x_{14}+x_{12}x_{14}+x_{13}x_{14}+x_2x_{15}+x_4x_{15}+x_6x_{15}+x_7x_{15}+x_9x_{15}+x_{11}x_{15}+x_{14}x_{15}+x_{15}+x_2x_{16}+x_4x_{16}+x_6x_{16}+x_7x_{16}+x_9x_{16}+x_{11}x_{16}+x_{14}x_{16}+x_2x_{17}+x_4x_{17}+x_6x_{17}+x_7x_{17}+x_9x_{17}+x_{11}x_{17}+x_{14}x_{17}+x_{17}+x_2x_{18}+x_4x_{18}+x_6x_{18}+x_7x_{18}+x_9x_{18}+x_{11}x_{18}+x_{14}x_{18}+x_2x_{19}+x_4x_{19}+x_6x_{19}+x_7x_{19}+x_9x_{19}+x_{11}x_{19}+x_{14}x_{19}+x_2x_{20}+x_3x_{20}+x_4x_{20}+x_6x_{20}+x_8x_{20}+x_9x_{20}+x_{11}x_{20}+x_{12}x_{20}+x_{13}x_{20}+x_{15}x_{20}+x_{16}x_{20}+x_{17}x_{20}+x_{18}x_{20}+x_{19}x_{20}+x_{20},$$

$Y_4 = (y_{41}, y_{42}, y_{43}, y_{44}, y_{45})$, where

$$y_{41} = x_4x_1+x_5x_1+x_7x_1+x_8x_1+x_9x_1+x_{10}x_1+x_{12}x_1+x_{13}x_1+x_{14}x_1+x_{15}x_1+x_{16}x_1+x_{17}x_1+x_1+x_3+x_4+x_3x_5+x_5+x_3x_6+x_5x_7+x_6x_7+x_2x_9+x_3x_9+x_4x_9+x_6x_9+x_8x_9+x_4x_{10}+x_5x_{10}+x_7x_{10}+x_8x_{10}+x_{10}+x_3x_{11}+x_4x_{11}+x_6x_{11}+x_8x_{11}+x_{10}x_{11}+x_2x_{12}+x_3x_{12}+x_5x_{12}+x_6x_{12}+x_7x_{12}+x_{10}x_{12}+x_2x_{13}+x_3x_{13}+x_4x_{13}+x_5x_{13}+x_8x_{13}+x_{11}x_{13}+x_{13}+x_3x_{14}+x_4x_{14}+x_5x_{14}+x_8x_{14}+x_9x_{14}+x_2x_{15}+x_4x_{15}+x_5x_{15}+x_7x_{15}+x_8x_{15}+x_{13}x_{15}+x_{15}+x_2x_{16}+x_5x_{16}+x_6x_{16}+x_{10}x_{16}+x_{11}x_{16}+x_{13}x_{16}+x_{15}x_{16}+x_{16}+x_3x_{17}+x_4x_{17}+x_6x_{17}+x_8x_{17}+x_{11}x_{17}+x_{12}x_{17}+x_{13}x_{17}+x_{14}x_{17}+x_2x_{18}+x_3x_{18}+x_4x_{18}+x_5x_{18}+x_8x_{18}+x_9x_{18}+x_{10}x_{18}+x_{13}x_{18}+x_{14}x_{18}+x_{16}x_{18}+x_2x_{19}+x_3x_{19}+x_5x_{19}+x_6x_{19}+x_7x_{19}+x_{11}x_{19}+x_{12}x_{19}+x_{14}x_{19}+x_{16}x_{19}+x_{19}+x_2x_{20}+x_4x_{20}+x_6x_{20}+x_7x_{20}+x_8x_{20}+x_{10}x_{20}+x_{12}x_{20}+x_{13}x_{20}+x_{15}x_{20}+x_{16}x_{20}+x_{19}x_{20},$$

$$y_{42} = x_5x_2+x_8x_2+x_9x_2+x_{10}x_2+x_{11}x_2+x_{13}x_2+x_{15}x_2+x_{16}x_2+x_{17}x_2+x_{18}x_2+x_{19}x_2+x_{20}x_2+x_2+x_4+x_1x_5+x_4x_5+x_3x_6+x_4x_6+x_5x_7+x_7+x_3x_8+x_5x_8+x_6x_8+x_7x_8+x_3x_9+x_7x_9+x_8x_9+x_1x_{10}+x_3x_{10}+x_5x_{10}+x_7x_{10}+x_{10}+x_1x_{11}+x_4x_{11}+x_5x_{11}+x_7x_{11}+x_8x_{11}+x_{12}+x_3x_{13}+x_5x_{13}+x_6x_{13}+x_7x_{13}+x_8x_{13}+x_{11}x_{13}+x_{13}+x_3x_{14}+x_4x_{14}+x_6x_{14}+x_8x_{14}+x_{10}x_{14}+x_{13}x_{14}+x_{14}+x_1x_{15}+x_4x_{15}+x_5x_{15}+x_7x_{15}+x_8x_{15}+x_{13}x_{15}+x_4x_{16}+x_5x_{16}+x_7x_{16}+x_{10}x_{16}+x_{11}x_{16}+x_{13}x_{16}+x_{14}x_{16}+x_{15}x_{16}+x_1x_{17}+x_4x_{17}+x_5x_{17}+x_6x_{17}+x_7x_{17}+x_9x_{17}+x_{10}x_{17}+x_{14}x_{17}+x_{16}x_{17}+x_3x_{18}+x_7x_{18}+x_8x_{18}+x_{17}x_{18}+x_{18}+x_3x_{19}+x_6x_{19}+x_7x_{19}+x_8x_{19}+x_{10}x_{19}+x_{14}x_{19}+x_{16}x_{19}+x_{17}x_{19}+x_{19}+x_4x_{20}+x_5x_{20}+x_6x_{20}+x_7x_{20}+x_{11}x_{20}+x_{13}x_{20}+x_{15}x_{20}+x_{16}x_{20}+x_{17}x_{20}+x_{19}x_{20}+x_{20},$$

$$y_{43} = 1+x_3x_2+x_4x_2+x_6x_2+x_9x_2+x_{10}x_2+x_{17}x_2+x_{18}x_2+x_{20}x_2+x_2+x_3x_4+x_4+x_1x_5+x_3x_5+x_4x_5+x_5+x_4x_6+x_6+x_3x_7+x_4x_7+x_6x_7+x_5x_8+x_6x_8+x_4x_9+x_5x_9+x_6x_9+x_7x_9+x_9+x_1x_{10}+x_3x_{10}+x_5x_{10}+x_7x_{10}+x_{10}+x_1x_{11}+x_3x_{11}+x_4x_{11}+x_5x_{11}+x_8x_{11}+x_9x_{11}+x_3x_{13}+x_4x_{13}+x_9x_{13}+x_3x_{14}+x_4x_{14}+x_6x_{14}+x_8x_{14}+x_{10}x_{14}+x_{13}x_{14}+x_1x_{15}+x_3x_{15}+x_4x_{15}+x_5x_{15}+x_8x_{15}+x_9x_{15}+x_{15}+x_6x_{16}+x_8x_{16}+x_9x_{16}+x_{10}x_{16}+x_{13}x_{16}+x_{14}x_{16}+x_{16}+x_1x_{17}+x_3x_{17}+x_7x_{17}+x_8x_{17}+x_9x_{17}+x_{10}x_{17}+x_{11}x_{17}+x_{13}x_{17}+x_{14}x_{17}+x_{15}x_{17}+x_4x_{18}+x_5x_{18}+x_6x_{18}+x_7x_{18}+x_{11}x_{18}+x_{13}x_{18}+x_{15}x_{18}+x_{16}x_{18}+x_{17}x_{18}+x_{18}+x_3x_{19}+x_4x_{19}+x_5x_{19}+x_9x_{19}+x_{10}x_{19}+x_{11}x_{19}+x_{14}x_{19}+x_{15}x_{19}+x_{16}x_{19}+x_{18}x_{19}+x_{19}+x_3x_{20}+x_7x_{20}+x_8x_{20}+x_{17}x_{20}+x_{20},$$

$$y_{44} = 1+x_5x_1+x_6x_1+x_8x_1+x_{11}x_1+x_{12}x_1+x_{14}x_1+x_{16}x_1+x_{17}x_1+x_{20}x_1+x_1+x_2+x_2x_3+x_2x_5+x_3x_5+x_5+x_2x_6+x_4x_6+x_6+x_2x_7+x_5x_7+x_6x_7+x_5x_8+x_4x_9+x_6x_9+x_7x_9+x_3x_{10}+x_4x_{10}+x_5x_{10}+x_6x_{10}+x_9x_{10}+x_5x_{11}+x_6x_{11}+x_9x_{11}+x_{10}x_{11}+x_3x_{12}+x_4x_{12}+x_5x_{12}+x_8x_{12}+x_9x_{12}+x_2x_{13}+x_3x_{13}+x_4x_{13}+x_{11}x_{13}+x_2x_{14}+x_4x_{14}+x_5x_{14}+x_7x_{14}+x_8x_{14}+x_{13}x_{14}+x_3x_{15}+x_4x_{15}+x_6x_{15}+x_8x_{15}+x_{10}x_{15}+x_{13}x_{15}+x_2x_{16}+x_5x_{16}+x_8x_{16}+x_9x_{16}+x_{11}x_{16}+x_{14}x_{16}+x_{15}x_{16}+x_{16}+x_3x_{17}+x_6x_{17}+x_7x_{17}+x_{12}x_{17}+x_{13}x_{17}+x_{15}x_{17}+x_{17}+x_2x_{18}+x_3x_{18}+x_4x_{18}+x_6x_{18}+x_9x_{18}+x_{10}x_{18}+x_{11}x_{18}+x_{12}x_{18}+x_{13}x_{18}+x_{17}x_{18}+x_{18}+x_4x_{19}+x_5x_{19}+x_6x_{19}+x_7x_{19}+x_8x_{19}+x_9x_{19}+x_{12}x_{19}+x_{13}x_{19}+x_{15}x_{19}+x_{17}x_{19}+x_{19}+x_2x_{20}+x_{10}x_{20}+x_{13}x_{20}+x_{14}x_{20}+x_{16}x_{20}+x_{17}x_{20}+x_{18}x_{20}+x_{19}x_{20}+x_{20},$$

$$y_{45} = 1+x_3x_2+x_5x_2+x_6x_2+x_7x_2+x_{13}x_2+x_{14}x_2+x_{16}x_2+x_{18}x_2+x_{20}x_2+x_2+x_1x_5+x_3x_5+x_5+x_1x_6+x_4x_6+x_5x_7+x_6x_7+x_1x_8+x_5x_8+x_8+x_4x_9+x_6x_9+x_7x_9+x_3x_{10}+x_4x_{10}+x_5x_{10}+x_6x_{10}+x_9x_{10}+x_1x_{11}+x_5x_{11}+x_6x_{11}+x_9x_{11}+x_{10}x_{11}+x_{11}+x_1x_{12}+x_3x_{12}+x_4x_{12}+x_5x_{12}+x_8x_{12}+x_9x_{12}+x_3x_{13}+x_4x_{13}+x_{11}x_{13}+x_1x_{14}+x_4x_{14}+x_5x_{14}+x_7x_{14}+x_8x_{14}+x_{13}x_{14}+x_3x_{15}+x_4x_{15}+x_6x_{15}+x_8x_{15}+x_{10}x_{15}+x_{13}x_{15}+x_{15}+x_1x_{16}+x_5x_{16}+x_8x_{16}+x_9x_{16}+x_{11}x_{16}+x_{14}x_{16}+x_{15}x_{16}+x_{16}+x_1x_{17}+x_3x_{17}+x_6x_{17}+x_7x_{17}+x_{12}x_{17}+x_{13}x_{17}+x_{15}x_{17}+x_3x_{18}+x_4x_{18}+x_6x_{18}+x_9x_{18}+x_{10}x_{18}+x_{11}x_{18}+x_{12}x_{18}+x_{13}x_{18}+x_{17}x_{18}+x_{18}+x_4x_{19}+x_5x_{19}+x_6x_{19}+x_7x_{19}+x_8x_{19}+x_9x_{19}+x_{12}x_{19}+x_{13}x_{19}+x_{15}x_{19}+x_{17}x_{19}+x_1x_{20}+x_{10}x_{20}+x_{13}x_{20}+x_{14}x_{20}+x_{16}x_{20}+x_{17}x_{20}+x_{18}x_{20}+x_{19}x_{20};$$

7) Set a 7-dimensional vector $Z$ that has all 7 coordinates as Boolean linear functions:

$Z = Y_1||Y_{2,1}||Y_{3,1} =$

$$\begin{bmatrix} x_4+x_5+x_7+x_{11}+x_{14}+x_{15}+x_{16}+x_{20} \\ \hline x_1+x_4+x_5+x_6+x_7+x_8+x_{10}+x_{11}+x_{12}+ \\ +x_{13}+x_{14}+x_{15}+x_{19}+x_{20} \\ \hline x_1+x_2+x_3+x_4+x_6+x_7+x_8+x_9+x_{10}+ \\ +x_{13}+x_{14}+x_{15}+x_{16}+x_{17}+x_{18}+x_{19} \\ \hline x_3+x_4+x_5+x_9+x_{10}+x_{11}+x_{12}+x_{14}+x_{17}+x_{19} \\ \hline x_1+x_2+x_3+x_7+x_8+x_{10}+x_{15}+x_{17}+x_{19}+x_{20} \\ \hline 1+x_1+x_4+x_7+x_8+x_{12}+x_{13}+x_{15}+x_{16}+x_{17} \\ \hline x_2+x_3+x_5+x_7+x_9+x_{10}+x_{12}+x_{13}+x_{15}+x_{18}+x_{20} \end{bmatrix}^T;$$

8) Transform $Z$ by the bijection of Dobbertin:

$W = \mathbf{Dob}(Z) = (W_1, W_2, W_3, W_4, W_5, W_6, W_7),$

where

$$W_1 = 1+x_2x_1+x_3x_1+x_5x_1+x_7x_1+x_{10}x_1+x_{11}x_1+x_{12}x_1+x_{13}x_1+x_{14}x_1+x_{15}x_1+x_{19}x_1+x_1+x_3+x_3x_4+x_4+x_2x_5+x_6+x_2x_7+x_3x_7+x_4x_7+x_5x_7+x_2x_8+x_3x_8+x_5x_8+x_7x_8+x_8+x_5x_9+x_6x_9+x_2x_{10}+x_4x_{10}+x_5x_{10}+x_7x_{10}+x_8x_{10}+x_9x_{10}+x_{10}+x_4x_{11}+x_5x_{11}+x_6x_{11}+x_7x_{11}+x_8x_{11}+x_9x_{11}+x_{10}x_{11}+x_2x_{12}+x_3x_{12}+x_4x_{12}+x_5x_{12}+x_8x_{12}+x_{10}x_{12}+x_{11}x_{12}+x_{12}+x_3x_{13}+x_5x_{13}+x_8x_{13}+x_{11}x_{13}+x_2x_{14}+x_4x_{14}+x_6x_{14}+x_7x_{14}+x_8x_{14}+x_9x_{14}+x_{10}x_{14}+x_{12}x_{14}+x_{14}+x_2x_{15}+x_3x_{15}+x_4x_{15}+x_5x_{15}+x_8x_{15}+x_{10}x_{15}+x_{11}x_{15}+x_{14}x_{15}+x_2x_{16}+x_3x_{16}+x_4x_{16}+x_7x_{16}+x_9x_{16}+x_{12}x_{16}+x_{14}x_{16}+x_{15}x_{16}+x_3x_{17}+x_4x_{17}+x_5x_{17}+x_7x_{17}+x_9x_{17}+x_{12}x_{17}+x_{13}x_{17}+x_{14}x_{17}+x_{15}x_{17}+x_{16}x_{17}+x_{17}+x_4x_{18}+x_6x_{18}+x_{11}x_{18}+x_{13}x_{18}+x_{16}x_{18}+x_{18}+x_6x_{19}+x_7x_{19}+x_8x_{19}+x_9x_{19}+x_{12}x_{19}+x_{14}x_{19}+x_{15}x_{19}+x_{17}x_{19}+x_{18}x_{19}+x_2x_{20}+x_6x_{20}+x_9x_{20}+x_{10}x_{20}+x_{13}x_{20}+x_{19}x_{20},$$

$$W_2 = 1+x_7x_1+x_{10}x_1+x_{11}x_1+x_{13}x_1+x_{15}x_1+x_{16}x_1+x_{17}x_1+x_1+x_2+x_2x_3+x_2x_4+x_4+x_2x_5+x_3x_5+x_5+x_4x_6+x_5x_6+x_3x_7+x_4x_7+x_5x_7+x_6x_7+x_7x_8+x_8+x_3x_9+x_4x_9+x_6x_9+x_7x_9+x_2x_{10}+x_3x_{10}+x_8x_{10}+x_9x_{10}+x_4x_{11}+x_7x_{11}+x_8x_{11}+x_{11}+x_2x_{12}+x_5x_{12}+x_9x_{12}+x_{11}x_{12}+x_{12}+x_5x_{13}+x_7x_{13}+x_8x_{13}+x_9x_{13}+x_{10}x_{13}+x_{11}x_{13}+x_{12}x_{13}+x_3x_{14}+x_7x_{14}+x_9x_{14}+x_{10}x_{14}+x_{12}x_{14}+x_3x_{15}+x_4x_{15}+x_5x_{15}+x_6x_{15}+x_8x_{15}+x_9x_{15}+x_{11}x_{15}+x_{13}x_{15}+x_{14}x_{15}+x_4x_{16}+x_5x_{16}+x_6x_{16}+x_7x_{16}+x_8x_{16}+x_{10}x_{16}+x_{11}x_{16}+x_{12}x_{16}+x_{13}x_{16}+x_{14}x_{16}+x_{15}x_{16}+x_2x_{17}+x_4x_{17}+x_5x_{17}+x_7x_{17}+x_8x_{17}+x_9x_{17}+x_{11}x_{17}+x_{12}x_{17}+x_{14}x_{17}+x_{15}x_{17}+x_2x_{18}+x_6x_{18}+x_9x_{18}+x_{10}x_{18}+x_{13}x_{18}+x_{18}+x_4x_{19}+x_6x_{19}+x_9x_{19}+x_{13}x_{19}+x_{14}x_{19}+x_{16}x_{19}+x_{18}x_{19}+x_2x_{20}+x_3x_{20}+x_5x_{20}+x_{12}x_{20}+x_{14}x_{20}+x_{16}x_{20}+x_{17}x_{20}+x_{19}x_{20}+x_{20},$$

$$W_3 = x_2x_1+x_4x_1+x_5x_1+x_6x_1+x_{11}x_1+x_{17}x_1+x_{18}x_1+x_1+x_3+x_2x_4+x_3x_4+x_4+x_2x_5+x_3x_5+x_2x_6+x_6+x_2x_7+x_4x_7+x_5x_7+x_6x_7+x_7+x_2x_8+x_4x_8+x_5x_8+x_6x_8+x_8+x_3x_9+x_7x_9+x_9+x_2x_{10}+x_3x_{10}+x_4x_{10}+x_5x_{10}+x_6x_{10}+x_9x_{10}+x_{10}+x_2x_{11}+x_3x_{11}+x_8x_{11}+x_9x_{11}+x_{11}+x_2x_{12}+x_6x_{12}+x_9x_{12}+x_{10}x_{12}+x_2x_{13}+x_3x_{13}+x_6x_{13}+x_{10}x_{13}+x_{11}x_{13}+x_{12}x_{13}+x_2x_{14}+x_6x_{14}+x_7x_{14}+x_9x_{14}+x_{11}x_{14}+x_{14}+x_2x_{15}+x_4x_{15}+x_5x_{15}+x_6x_{15}+x_9x_{15}+x_{14}x_{15}+x_{15}+x_9x_{16}+x_{11}x_{16}+x_{14}x_{16}+x_{16}+x_3x_{17}+x_4x_{17}+x_5x_{17}+x_6x_{17}+x_7x_{17}+x_8x_{17}+x_9x_{17}+x_{10}x_{17}+x_{14}x_{17}+x_{15}x_{17}+x_3x_{18}+x_6x_{18}+x_8x_{18}+x_9x_{18}+x_{10}x_{18}+x_{11}x_{18}+x_{16}x_{18}+x_{18}+x_2x_{19}+x_3x_{19}+x_4x_{19}+x_5x_{19}+x_6x_{19}+x_7x_{19}+x_{10}x_{19}+x_{12}x_{19}+x_{14}x_{19}+x_{15}x_{19}+x_{16}x_{19}+x_{19}+x_2x_{20}+x_4x_{20}+x_5x_{20}+x_6x_{20}+x_7x_{20}+x_{10}x_{20}+x_{13}x_{20}+x_{15}x_{20}+x_{16}x_{20}+x_{18}x_{20},$$

$$W_4 = x_7x_1+x_9x_1+x_{10}x_1+x_{13}x_1+x_{14}x_1+x_{15}x_1+x_{16}x_1+x_{17}x_1+x_{18}x_1+x_{19}x_1+x_{20}x_1+x_1+x_2x_3+x_3+x_2x_4+x_2x_5+x_3x_5+x_5+x_4x_6+x_5x_6+x_2x_7+x_3x_7+x_5x_7+x_6x_7+x_7+x_7x_8+x_8+x_3x_9+x_5x_9+x_7x_9+x_8x_9+x_3x_{10}+x_4x_{10}+x_7x_{10}+x_8x_{10}+x_9x_{10}+x_{10}+x_2x_{11}+x_4x_{11}+x_5x_{11}+x_6x_{11}+x_7x_{11}+x_{10}x_{11}+x_2x_{12}+x_5x_{12}+x_2x_{13}+x_4x_{13}+x_5x_{13}+x_7x_{13}+x_8x_{13}+x_9x_{13}+x_{11}x_{13}+x_{12}x_{13}+x_3x_{14}+x_4x_{14}+x_5x_{14}+x_6x_{14}+x_8x_{14}+x_{13}x_{14}+x_{14}+x_2x_{15}+x_3x_{15}+x_5x_{15}+x_6x_{15}+x_8x_{15}+x_9x_{15}+x_{10}x_{15}+x_{11}x_{15}+x_{13}x_{15}+x_{15}+x_2x_{16}+x_5x_{16}+x_6x_{16}+x_8x_{16}+x_9x_{16}+x_{10}x_{16}+x_{11}x_{16}+x_{12}x_{16}+x_{14}x_{16}+x_{16}+x_5x_{17}+x_7x_{17}+x_8x_{17}+x_{10}x_{17}+x_{12}x_{17}+x_{14}x_{17}+x_{15}x_{17}+x_{16}x_{17}+x_{17}+x_2x_{18}+x_4x_{18}+x_5x_{18}+x_7x_{18}+x_8x_{18}+x_9x_{18}+x_{11}x_{18}+x_{12}x_{18}+x_{14}x_{18}+x_{15}x_{18}+x_{18}+x_5x_{19}+x_7x_{19}+x_8x_{19}+x_{10}x_{19}+x_{12}x_{19}+x_{14}x_{19}+x_{15}x_{19}+x_{16}x_{19}+x_2x_{20}+x_3x_{20}+x_4x_{20}+x_6x_{20}+x_7x_{20}+x_8x_{20}+x_9x_{20}+x_{10}x_{20}+x_{13}x_{20}+x_{14}x_{20}+x_{15}x_{20}+x_{16}x_{20}+x_{17}x_{20}+x_{18}x_{20}+x_{19}x_{20}+x_{20},$$

$$W_5 = x_4x_2+x_6x_2+x_{11}x_2+x_{13}x_2+x_{16}x_2+x_{19}x_2+x_2+x_3+x_1x_4+x_4+x_1x_5+x_5x_6+x_6+x_1x_7+x_6x_7+x_7+x_4x_8+x_5x_8+x_7x_8+x_1x_9+x_5x_9+x_7x_9+x_8x_9+x_9+x_5x_{10}+x_6x_{10}+x_7x_{10}+x_9x_{10}+x_4x_{11}+x_5x_{11}+x_6x_{11}+x_7x_{11}+x_{10}x_{11}+x_{11}+x_4x_{12}+x_5x_{12}+x_7x_{12}+$$

$x_9x_{12} + x_{12} + x_5x_{13} + x_7x_{13} + x_{10}x_{13} + x_{11}x_{13} + x_{13} + x_4x_{14} + x_5x_{14} + x_7x_{14} + x_9x_{14} + x_1x_{15} + x_6x_{15} + x_8x_{15} + x_9x_{15} + x_{10}x_{15} + x_{11}x_{15} +$
$x_{12}x_{15} + x_{13}x_{15} + x_{14}x_{15} + x_{15} + x_1x_{16} + x_4x_{16} + x_5x_{16} + x_7x_{16} + x_8x_{16} + x_9x_{16} + x_{11}x_{16} + x_{12}x_{16} + x_{14}x_{16} + x_{15}x_{16} + x_{16} + x_{17} + x_1x_{18} +$
$x_4x_{18} + x_5x_{18} + x_6x_{18} + x_7x_{18} + x_8x_{18} + x_{10}x_{18} + x_{11}x_{18} + x_{12}x_{18} + x_{13}x_{18} + x_{14}x_{18} + x_{15}x_{18} + x_{18} + x_1x_{19} + x_4x_{19} + x_6x_{19} + x_8x_{19} +$
$x_{12}x_{19} + x_{13}x_{19} + x_{14}x_{19} + x_{18}x_{19} + x_4x_{20} + x_5x_{20} + x_7x_{20} + x_9x_{20} + x_{15}x_{20} + x_{16}x_{20} + x_{18}x_{20} + x_{19}x_{20} + x_{20},$

$W_6 = 1 + x_1x_2 + x_5x_2 + x_8x_2 + x_{11}x_2 + x_{12}x_2 + x_{13}x_2 + x_{14}x_2 + x_{17}x_2 + x_{20}x_2 + x_1x_3 + x_3 + x_1x_4 + x_3x_4 + x_4 + x_4x_5 + x_5 + x_1x_7 +$
$x_3x_7 + x_4x_7 + x_5x_7 + x_6x_7 + x_7 + x_3x_8 + x_4x_8 + x_7x_8 + x_3x_9 + x_4x_9 + x_5x_9 + x_4x_{10} + x_6x_{10} + x_9x_{10} + x_1x_{11} + x_3x_{11} + x_4x_{11} + x_6x_{11} +$
$x_8x_{11} + x_{11} + x_1x_{12} + x_3x_{12} + x_8x_{12} + x_9x_{12} + x_3x_{13} + x_4x_{13} + x_5x_{13} + x_6x_{13} + x_7x_{13} + x_{10}x_{13} + x_{12}x_{13} + x_3x_{14} + x_4x_{14} + x_{10}x_{14} + x_{12}x_{14} +$
$x_{13}x_{14} + x_{14} + x_1x_{15} + x_3x_{15} + x_4x_{15} + x_5x_{15} + x_6x_{15} + x_8x_{15} + x_{13}x_{15} + x_{15} + x_3x_{16} + x_4x_{16} + x_6x_{16} + x_7x_{16} + x_{10}x_{16} + x_{12}x_{16} + x_{13}x_{16} +$
$x_{14}x_{16} + x_{15}x_{16} + x_1x_{17} + x_3x_{17} + x_5x_{17} + x_6x_{17} + x_7x_{17} + x_8x_{17} + x_9x_{17} + x_{15}x_{17} + x_{16}x_{17} + x_{17} + x_3x_{18} + x_7x_{18} + x_9x_{18} + x_{10}x_{18} +$
$x_{12}x_{18} + x_{15}x_{18} + x_{16}x_{18} + x_{17}x_{18} + x_1x_{19} + x_3x_{19} + x_4x_{19} + x_7x_{19} + x_8x_{19} + x_{14}x_{19} + x_{15}x_{19} + x_{16}x_{19} + x_{17}x_{19} + x_{18}x_{19} + x_1x_{20} + x_3x_{20} +$
$x_7x_{20} + x_8x_{20} + x_9x_{20} + x_{10}x_{20} + x_{11}x_{20} + x_{14}x_{20} + x_{15}x_{20} + x_{17}x_{20} + x_{18}x_{20} + x_{20},$

$W_7 = x_1x_2 + x_3x_2 + x_5x_2 + x_7x_2 + x_8x_2 + x_{10}x_2 + x_{13}x_2 + x_{15}x_2 + x_{16}x_2 + x_{18}x_2 + x_{20}x_2 + x_3 + x_3x_5 + x_1x_6 + x_5x_6 + x_1x_7 + x_4x_7 +$
$x_5x_7 + x_6x_8 + x_7x_8 + x_1x_9 + x_4x_9 + x_7x_9 + x_8x_9 + x_9 + x_4x_{10} + x_7x_{10} + x_{10} + x_1x_{11} + x_6x_{11} + x_8x_{11} + x_{10}x_{11} + x_{11} + x_5x_{12} + x_6x_{12} + x_7x_{12} +$
$x_9x_{12} + x_{11}x_{12} + x_{12} + x_3x_{13} + x_4x_{13} + x_5x_{13} + x_6x_{13} + x_7x_{13} + x_9x_{13} + x_{10}x_{13} + x_{11}x_{13} + x_{12}x_{13} + x_4x_{14} + x_6x_{14} + x_7x_{14} + x_{10}x_{14} + x_{14} +$
$x_1x_{15} + x_4x_{15} + x_5x_{15} + x_8x_{15} + x_9x_{15} + x_{10}x_{15} + x_{12}x_{15} + x_{13}x_{15} + x_{14}x_{15} + x_1x_{16} + x_3x_{16} + x_4x_{16} + x_5x_{16} + x_7x_{16} + x_8x_{16} + x_9x_{16} +$
$x_{11}x_{16} + x_{13}x_{16} + x_{15}x_{16} + x_{16} + x_1x_{17} + x_3x_{17} + x_4x_{17} + x_5x_{17} + x_6x_{17} + x_7x_{17} + x_8x_{17} + x_9x_{17} + x_{10}x_{17} + x_{14}x_{17} + x_{15}x_{17} + x_{17} + x_3x_{18} +$
$x_4x_{18} + x_{10}x_{18} + x_{12}x_{18} + x_{13}x_{18} + x_{16}x_{18} + x_{18} + x_1x_{19} + x_3x_{19} + x_4x_{19} + x_8x_{19} + x_{10}x_{19} + x_{11}x_{19} + x_{13}x_{19} + x_{14}x_{19} + x_{18}x_{19} + x_{19} +$
$x_4x_{20} + x_5x_{20} + x_6x_{20} + x_7x_{20} + x_{10}x_{20} + x_{13}x_{20} + x_{15}x_{20} + x_{16}x_{20} + x_{18}x_{20};$

9) Set $Y_1 = (W_1, W_2, W_3, W_4, W_5)$, $Y_{2,1} = W_6$, $Y_{3,1} = W_7$;

10) Compute $y = T \cdot y'^T$;

11) The public key is $y$, given by the system of 20 equations of 20 unknowns:

$$\begin{aligned} &y_1 = P_1(x_1, x_2, \dots, x_{20}),\\ &y_2 = P_2(x_1, x_2, \dots, x_{20}),\\ &\dots\\ &y_{20} = P_{20}(x_1, x_2, \dots, x_{20}), \end{aligned}$$

where $P_i$ are multivariate quadratic polynomials of 20 Boolean variables.

In developed form, they look like these:

$y_1 = P_1(x) = x_1 + x_2 + x_2x_1 + x_3x_1 + x_4x_1 + x_5x_1 + x_6x_1 + x_7x_1 + x_8x_1 + x_{10}x_1 + x_3x_4 + x_4x_5 + x_5 + x_4x_6 + x_5x_6 + x_6 + x_7 + x_2x_8 + x_3x_8 + x_4x_8 +$
$x_5x_8 + x_7x_8 + x_8 + x_2x_9 + x_3x_9 + x_7x_9 + x_8x_9 + x_9 + x_2x_{10} + x_4x_{10} + x_5x_{10} + x_6x_{10} + x_7x_{10} + x_9x_{10} + x_2x_{11} + x_4x_{11} + x_7x_{11} + x_8x_{11} + x_9x_{11} + x_2x_{12} +$
$x_5x_{12} + x_7x_{12} + x_8x_{12} + x_9x_{12} + x_{11}x_{12} + x_2x_{13} + x_4x_{13} + x_8x_{13} + x_9x_{13} + x_{10}x_{13} + x_2x_{14} + x_3x_{14} + x_9x_{14} + x_{10}x_{14} + x_{11}x_{14} + x_{13}x_{14} + x_2x_{15} +$
$x_4x_{15} + x_5x_{15} + x_7x_{15} + x_9x_{15} + x_{11}x_{15} + x_{14}x_{15} + x_{15} + x_4x_{16} + x_5x_{16} + x_6x_{16} + x_7x_{16} + x_9x_{16} + x_{10}x_{16} + x_{11}x_{16} + x_{15}x_{16} + x_3x_{17} + x_4x_{17} + x_6x_{17} +$
$x_7x_{17} + x_8x_{17} + x_{12}x_{17} + x_{16}x_{17} + x_{17} + x_4x_{18} + x_6x_{18} + x_7x_{18} + x_{12}x_{18} + x_{13}x_{18} + x_{15}x_{18} + x_{16}x_{18} + x_{18} + x_2x_{19} + x_5x_{19} + x_6x_{19} + x_7x_{19} + x_8x_{19} +$
$x_9x_{19} + x_{11}x_{19} + x_{13}x_{19} + x_{14}x_{19} + x_{16}x_{19} + x_{17}x_{19} + x_{19} + x_3x_{20} + x_5x_{20} + x_6x_{20} + x_8x_{20} + x_9x_{20} + x_{13}x_{20} + x_{15}x_{20} + x_{16}x_{20} + x_{17}x_{20} + x_{19}x_{20},$

$y_2 = P_2(x) = x_2x_1 + x_3x_1 + x_4x_1 + x_8x_1 + x_{10}x_1 + x_{12}x_1 + x_{17}x_1 + x_{19}x_1 + x_{20}x_1 + x_1 + x_2 + x_2x_3 + x_3 + x_3x_4 + x_4 + x_2x_5 + x_2x_6 +$
$x_5x_6 + x_6 + x_2x_7 + x_3x_7 + x_7 + x_2x_8 + x_3x_8 + x_4x_8 + x_7x_8 + x_8 + x_5x_9 + x_7x_9 + x_9 + x_3x_{10} + x_4x_{10} + x_6x_{10} + x_7x_{10} + x_4x_{11} + x_5x_{11} + x_6x_{11} +$
$x_7x_{11} + x_8x_{11} + x_9x_{11} + x_{10}x_{11} + x_4x_{12} + x_5x_{12} + x_{10}x_{12} + x_3x_{13} + x_5x_{13} + x_7x_{13} + x_{11}x_{13} + x_{12}x_{13} + x_{13} + x_2x_{14} + x_7x_{14} + x_{11}x_{14} +$
$x_{13}x_{14} + x_{14} + x_2x_{15} + x_5x_{15} + x_{12}x_{15} + x_{13}x_{15} + x_{15} + x_2x_{16} + x_3x_{16} + x_4x_{16} + x_5x_{16} + x_6x_{16} + x_9x_{16} + x_{12}x_{16} + x_{13}x_{16} + x_{14}x_{16} + x_3x_{17} +$
$x_4x_{17} + x_6x_{17} + x_7x_{17} + x_9x_{17} + x_2x_{18} + x_4x_{18} + x_7x_{18} + x_9x_{18} + x_{12}x_{18} + x_{13}x_{18} + x_{14}x_{18} + x_{15}x_{18} + x_{17}x_{18} + x_4x_{19} + x_6x_{19} + x_7x_{19} +$
$x_8x_{19} + x_{11}x_{19} + x_{18}x_{19} + x_{19} + x_2x_{20} + x_3x_{20} + x_6x_{20} + x_7x_{20} + x_{15}x_{20} + x_{17}x_{20} + x_{19}x_{20},$

$y_3 = P_3(x) = x_2x_1 + x_3x_1 + x_9x_1 + x_{13}x_1 + x_{16}x_1 + x_{17}x_1 + x_{19}x_1 + x_{20}x_1 + x_1 + x_3 + x_2x_4 + x_3x_4 + x_2x_5 + x_5 + x_2x_6 + x_3x_7 + x_4x_7 + x_5x_7 +$
$x_6x_7 + x_7 + x_2x_8 + x_3x_8 + x_3x_9 + x_4x_9 + x_5x_9 + x_6x_9 + x_9 + x_2x_{10} + x_3x_{10} + x_4x_{10} + x_6x_{10} + x_8x_{10} + x_{10} + x_2x_{11} + x_4x_{11} + x_6x_{11} + x_2x_{12} + x_3x_{12} +$
$x_4x_{12} + x_{11}x_{12} + x_3x_{13} + x_6x_{13} + x_7x_{13} + x_8x_{13} + x_9x_{13} + x_5x_{14} + x_6x_{14} + x_7x_{14} + x_8x_{14} + x_{12}x_{14} + x_{13}x_{14} + x_2x_{15} + x_4x_{15} + x_5x_{15} + x_6x_{15} +$
$x_{11}x_{15} + x_{12}x_{15} + x_{13}x_{15} + x_{15} + x_3x_{16} + x_4x_{16} + x_5x_{16} + x_6x_{16} + x_9x_{16} + x_{10}x_{16} + x_{12}x_{16} + x_{14}x_{16} + x_{16} + x_4x_{17} + x_5x_{17} + x_6x_{17} + x_7x_{17} +$
$x_8x_{17} + x_9x_{17} + x_{10}x_{17} + x_{12}x_{17} + x_{13}x_{17} + x_4x_{18} + x_6x_{18} + x_8x_{18} + x_{10}x_{18} + x_{12}x_{18} + x_{13}x_{18} + x_{14}x_{18} + x_{15}x_{18} + x_{17}x_{18} + x_{18} + x_3x_{19} + x_4x_{19} +$
$x_7x_{19} + x_{13}x_{19} + x_{17}x_{19} + x_3x_{20} + x_4x_{20} + x_5x_{20} + x_6x_{20} + x_7x_{20} + x_8x_{20} + x_9x_{20} + x_{10}x_{20} + x_{12}x_{20} + x_{13}x_{20} + x_{14}x_{20} + x_{16}x_{20} + x_{17}x_{20} + x_{18}x_{20},$

$y_4 = P_4(x) = x_4x_1 + x_7x_1 + x_8x_1 + x_{10}x_1 + x_{11}x_1 + x_{12}x_1 + x_{13}x_1 + x_{18}x_1 + x_{19}x_1 + x_{20}x_1 + x_1 + x_2 + x_2x_5 + x_4x_5 + x_5 + x_2x_6 + x_3x_6 +$
$x_4x_6 + x_6 + x_3x_7 + x_6x_7 + x_7 + x_2x_8 + x_4x_8 + x_5x_8 + x_6x_8 + x_7x_8 + x_3x_9 + x_5x_9 + x_7x_9 + x_2x_{10} + x_4x_{10} + x_6x_{10} + x_7x_{10} + x_2x_{11} + x_4x_{11} + x_7x_{11} +$
$x_8x_{11} + x_9x_{11} + x_{11} + x_2x_{12} + x_3x_{12} + x_4x_{12} + x_8x_{12} + x_9x_{12} + x_{10}x_{12} + x_{11}x_{12} + x_3x_{13} + x_5x_{13} + x_6x_{13} + x_7x_{13} + x_9x_{13} + x_{11}x_{13} + x_{13} +$
$x_3x_{14} + x_4x_{14} + x_5x_{14} + x_6x_{14} + x_{10}x_{14} + x_{11}x_{14} + x_{13}x_{14} + x_{14} + x_2x_{15} + x_3x_{15} + x_4x_{15} + x_5x_{15} + x_7x_{15} + x_9x_{15} + x_{10}x_{15} + x_{13}x_{15} + x_2x_{16} +$
$x_3x_{16} + x_4x_{16} + x_6x_{16} + x_9x_{16} + x_{11}x_{16} + x_{12}x_{16} + x_{15}x_{16} + x_2x_{17} + x_3x_{17} + x_4x_{17} + x_7x_{17} + x_8x_{17} + x_{10}x_{17} + x_{13}x_{17} + x_{14}x_{17} + x_{15}x_{17} +$
$x_{17} + x_2x_{18} + x_3x_{18} + x_6x_{18} + x_7x_{18} + x_8x_{18} + x_9x_{18} + x_{10}x_{18} + x_{12}x_{18} + x_{14}x_{18} + x_{15}x_{18} + x_2x_{19} + x_5x_{19} + x_7x_{19} + x_8x_{19} + x_9x_{19} + x_{11}x_{19} +$
$x_{14}x_{19} + x_{16}x_{19} + x_{17}x_{19} + x_{18}x_{19} + x_2x_{20} + x_3x_{20} + x_5x_{20} + x_7x_{20} + x_8x_{20} + x_9x_{20} + x_{11}x_{20} + x_{12}x_{20} + x_{14}x_{20} + x_{16}x_{20} + x_{19}x_{20} + x_{20} + 1,$

$y_5 = P_5(x) = x_2x_1 + x_4x_1 + x_9x_1 + x_{10}x_1 + x_{13}x_1 + x_{17}x_1 + x_{19}x_1 + x_1 + x_2 + x_2x_3 + x_3 + x_2x_4 + x_3x_4 + x_2x_5 + x_4x_5 + x_4x_6 + x_5x_6 + x_2x_7 +$
$x_5x_7 + x_6x_7 + x_5x_8 + x_6x_8 + x_7x_8 + x_8 + x_2x_9 + x_3x_9 + x_4x_9 + x_5x_9 + x_7x_9 + x_8x_9 + x_9 + x_3x_{10} + x_5x_{10} + x_8x_{10} + x_9x_{10} + x_{10} + x_4x_{11} + x_5x_{11} + x_6x_{11} +$
$x_8x_{11} + x_{11} + x_3x_{12} + x_5x_{12} + x_8x_{12} + x_9x_{12} + x_{12} + x_2x_{13} + x_4x_{13} + x_6x_{13} + x_7x_{13} + x_2x_{14} + x_3x_{14} + x_4x_{14} + x_6x_{14} + x_7x_{14} + x_9x_{14} + x_{10}x_{14} +$
$x_2x_{15} + x_3x_{15} + x_5x_{15} + x_6x_{15} + x_{11}x_{15} + x_{12}x_{15} + x_{14}x_{15} + x_2x_{16} + x_3x_{16} + x_5x_{16} + x_6x_{16} + x_7x_{16} + x_{11}x_{16} + x_{12}x_{16} + x_{15}x_{16} + x_2x_{17} + x_3x_{17} +$
$x_5x_{17} + x_6x_{17} + x_8x_{17} + x_{10}x_{17} + x_{11}x_{17} + x_{16}x_{17} + x_{17} + x_2x_{18} + x_3x_{18} + x_5x_{18} + x_6x_{18} + x_7x_{18} + x_8x_{18} + x_9x_{18} + x_{11}x_{18} + x_{12}x_{18} + x_{13}x_{18} +$
$x_{14}x_{18} + x_{16}x_{18} + x_3x_{19} + x_5x_{19} + x_7x_{19} + x_8x_{19} + x_{12}x_{19} + x_{13}x_{19} + x_{16}x_{19} + x_{18}x_{19} + x_{19} + x_6x_{20} + x_{11}x_{20} + x_{12}x_{20} + x_{13}x_{20} + x_{17}x_{20} + x_{18}x_{20} + 1,$

$y_6 = P_6(x) = x_1x_2 + x_3x_2 + x_5x_2 + x_6x_2 + x_7x_2 + x_8x_2 + x_{15}x_2 + x_{17}x_2 + x_{18}x_2 + x_{20}x_2 + x_1x_3 + x_3 + x_5 + x_1x_6 + x_3x_6 + x_4x_6 + x_1x_7 +$
$x_4x_7 + x_5x_7 + x_1x_8 + x_3x_8 + x_7x_8 + x_8 + x_4x_9 + x_8x_9 + x_4x_{10} + x_6x_{10} + x_8x_{10} + x_{10} + x_1x_{11} + x_3x_{11} + x_9x_{11} + x_{10}x_{11} + x_{11} + x_1x_{12} + x_5x_{12} +$
$x_7x_{12} + x_8x_{12} + x_{11}x_{12} + x_1x_{13} + x_3x_{13} + x_4x_{13} + x_6x_{13} + x_8x_{13} + x_9x_{13} + x_{12}x_{13} + x_{13} + x_3x_{14} + x_5x_{14} + x_6x_{14} + x_8x_{14} + x_{10}x_{14} + x_{11}x_{14} +$
$x_{13}x_{14} + x_{14} + x_3x_{15} + x_7x_{15} + x_8x_{15} + x_9x_{15} + x_{10}x_{15} + x_{11}x_{15} + x_{12}x_{15} + x_{14}x_{15} + x_7x_{16} + x_8x_{16} + x_9x_{16} + x_{10}x_{16} + x_{13}x_{16} + x_{15}x_{16} +$
$x_1x_{17} + x_4x_{17} + x_9x_{17} + x_{11}x_{17} + x_{17} + x_1x_{18} + x_4x_{18} + x_5x_{18} + x_{11}x_{18} + x_{14}x_{18} + x_{16}x_{18} + x_1x_{19} + x_4x_{19} + x_5x_{19} + x_6x_{19} + x_7x_{19} + x_8x_{19} +$
$x_{10}x_{19} + x_{14}x_{19} + x_{15}x_{19} + x_{19} + x_1x_{20} + x_5x_{20} + x_6x_{20} + x_7x_{20} + x_9x_{20} + x_{10}x_{20} + x_{11}x_{20} + x_{15}x_{20} + x_{16}x_{20} + x_{17}x_{20} + x_{18}x_{20} + x_{19}x_{20} + 1,$

$y_7 = P_7(x) = x_4x_3 + x_5x_3 + x_6x_3 + x_{13}x_3 + x_{14}x_3 + x_{16}x_3 + x_{18}x_3 + x_{20}x_3 + x_3 + x_4 + x_1x_5 + x_2x_5 + x_4x_5 + x_1x_6 + x_2x_6 + x_6 + x_1x_7 + x_5x_7 +$
$x_1x_8 + x_7x_8 + x_1x_9 + x_2x_9 + x_4x_9 + x_5x_9 + x_1x_{10} + x_2x_{10} + x_5x_{10} + x_6x_{10} + x_8x_{10} + x_9x_{10} + x_5x_{11} + x_7x_{11} + x_8x_{11} + x_9x_{11} + x_{11} + x_1x_{12} + x_2x_{12} +$
$x_4x_{12} + x_6x_{12} + x_9x_{12} + x_{11}x_{12} + x_1x_{13} + x_4x_{13} + x_5x_{13} + x_6x_{13} + x_7x_{13} + x_9x_{13} + x_{11}x_{13} + x_{12}x_{13} + x_1x_{14} + x_2x_{14} + x_4x_{14} + x_7x_{14} + x_8x_{14} +$
$x_9x_{14} + x_{13}x_{14} + x_{14} + x_4x_{15} + x_7x_{15} + x_9x_{15} + x_{10}x_{15} + x_{14}x_{15} + x_4x_{16} + x_6x_{16} + x_7x_{16} + x_8x_{16} + x_9x_{16} + x_{13}x_{16} + x_{14}x_{16} + x_{16} + x_2x_{17} + x_4x_{17} +$
$x_5x_{17} + x_7x_{17} + x_{13}x_{17} + x_{14}x_{17} + x_{15}x_{17} + x_{16}x_{17} + x_{17} + x_1x_{18} + x_5x_{18} + x_6x_{18} + x_8x_{18} + x_{12}x_{18} + x_{15}x_{18} + x_{18} + x_4x_{19} + x_5x_{19} + x_7x_{19} +$
$x_8x_{19} + x_9x_{19} + x_{14}x_{19} + x_{15}x_{19} + x_{18}x_{19} + x_4x_{20} + x_5x_{20} + x_6x_{20} + x_7x_{20} + x_9x_{20} + x_{11}x_{20} + x_{13}x_{20} + x_{14}x_{20} + x_{16}x_{20} + x_{17}x_{20} + x_{18}x_{20},$

$y_8 = P_8(x) = x_4x_1 + x_5x_1 + x_7x_1 + x_8x_1 + x_{14}x_1 + x_{15}x_1 + x_{16}x_1 + x_{19}x_1 + x_{20}x_1 + x_1 + x_2 + x_2x_3 + x_3 + x_2x_4 + x_4 + x_3x_5 + x_4x_5 + x_5 +$
$x_2x_6 + x_3x_6 + x_2x_7 + x_3x_7 + x_5x_7 + x_7 + x_4x_8 + x_5x_8 + x_6x_8 + x_2x_9 + x_5x_9 + x_9 + x_2x_{10} + x_3x_{10} + x_5x_{10} + x_9x_{10} + x_{10} + x_4x_{11} + x_5x_{11} + x_6x_{11} +$
$x_7x_{11} + x_8x_{11} + x_9x_{11} + x_{10}x_{11} + x_{11} + x_2x_{12} + x_3x_{12} + x_5x_{12} + x_7x_{12} + x_8x_{12} + x_9x_{12} + x_{11}x_{12} + x_{12} + x_5x_{13} + x_7x_{13} + x_{12}x_{13} + x_2x_{14} + x_3x_{14} +$
$x_6x_{14} + x_9x_{14} + x_{10}x_{14} + x_2x_{15} + x_3x_{15} + x_5x_{15} + x_6x_{15} + x_7x_{15} + x_8x_{15} + x_9x_{15} + x_{12}x_{15} + x_3x_{16} + x_4x_{16} + x_9x_{16} + x_{13}x_{16} + x_{15}x_{16} + x_5x_{17} +$
$x_6x_{17} + x_8x_{17} + x_9x_{17} + x_{10}x_{17} + x_{14}x_{17} + x_2x_{18} + x_3x_{18} + x_4x_{18} + x_5x_{18} + x_7x_{18} + x_9x_{18} + x_{11}x_{18} + x_{13}x_{18} + x_{16}x_{18} + x_{18} + x_4x_{19} + x_8x_{19} +$
$x_9x_{19} + x_{10}x_{19} + x_{12}x_{19} + x_{15}x_{19} + x_{17}x_{19} + x_{18}x_{19} + x_6x_{20} + x_7x_{20} + x_8x_{20} + x_{10}x_{20} + x_{11}x_{20} + x_{12}x_{20} + x_{14}x_{20} + x_{15}x_{20} + x_{19}x_{20} + x_{20},$

$y_9 = P_9(x) = x_2x_1 + x_4x_1 + x_9x_1 + x_{11}x_1 + x_{13}x_1 + x_{16}x_1 + x_{17}x_1 + x_{19}x_1 + x_1 + x_2x_3 + x_2x_4 + x_3x_4 + x_2x_5 + x_4x_5 + x_3x_6 + x_5x_6 +$
$x_3x_7 + x_5x_7 + x_6x_7 + x_2x_8 + x_5x_8 + x_6x_8 + x_8 + x_2x_9 + x_5x_9 + x_8x_9 + x_9 + x_4x_{10} + x_5x_{10} + x_7x_{10} + x_9x_{10} + x_2x_{11} + x_5x_{11} + x_6x_{11} + x_8x_{11} +$
$x_9x_{11} + x_{11} + x_2x_{12} + x_5x_{12} + x_6x_{12} + x_8x_{12} + x_{11}x_{12} + x_{12} + x_3x_{13} + x_6x_{13} + x_9x_{13} + x_{10}x_{13} + x_{13} + x_4x_{14} + x_6x_{14} + x_{10}x_{14} + x_2x_{15} +$
$x_4x_{15} + x_6x_{15} + x_7x_{15} + x_8x_{15} + x_{10}x_{15} + x_{11}x_{15} + x_2x_{16} + x_3x_{16} + x_5x_{16} + x_{10}x_{16} + x_{12}x_{16} + x_4x_{17} + x_6x_{17} + x_9x_{17} + x_{10}x_{17} + x_{11}x_{17} +$
$x_{12}x_{17} + x_{15}x_{17} + x_{17} + x_4x_{18} + x_7x_{18} + x_9x_{18} + x_{11}x_{18} + x_{13}x_{18} + x_{15}x_{18} + x_{16}x_{18} + x_{17}x_{18} + x_{18} + x_3x_{19} + x_4x_{19} + x_{12}x_{19} + x_{16}x_{19} +$
$x_{17}x_{19} + x_3x_{20} + x_4x_{20} + x_5x_{20} + x_6x_{20} + x_8x_{20} + x_9x_{20} + x_{10}x_{20} + x_{11}x_{20} + x_{13}x_{20} + x_{14}x_{20} + x_{15}x_{20} + x_{16}x_{20} + x_{20},$

$$y_{10} = P_{10}(x) = x_1x_2 + x_4x_2 + x_5x_2 + x_6x_2 + x_7x_2 + x_8x_2 + x_9x_2 + x_{10}x_2 + x_{11}x_2 + x_{12}x_2 + x_{15}x_2 + x_{16}x_2 + x_{18}x_2 + x_2 + x_1x_3 + x_4 + x_1x_5 + x_3x_5 + x_4x_5 + x_1x_6 + x_5x_6 + x_3x_7 + x_5x_7 + x_6x_7 + x_7 + x_4x_8 + x_6x_8 + x_8 + x_1x_9 + x_4x_9 + x_7x_9 + x_3x_{10} + x_4x_{10} + x_7x_{10} + x_8x_{10} + x_9x_{10} + x_1x_{11} + x_4x_{11} + x_5x_{11} + x_{11} + x_5x_{12} + x_6x_{12} + x_7x_{12} + x_9x_{12} + x_{12} + x_1x_{13} + x_5x_{13} + x_7x_{13} + x_{13} + x_4x_{14} + x_5x_{14} + x_7x_{14} + x_{11}x_{14} + x_{14} + x_1x_{15} + x_3x_{15} + x_4x_{15} + x_5x_{15} + x_6x_{15} + x_7x_{15} + x_{11}x_{15} + x_{12}x_{15} + x_{14}x_{15} + x_1x_{16} + x_4x_{16} + x_7x_{16} + x_{10}x_{16} + x_{11}x_{16} + x_{12}x_{16} + x_{13}x_{16} + x_{15}x_{16} + x_{16} + x_8x_{17} + x_9x_{17} + x_{11}x_{17} + x_{13}x_{17} + x_{14}x_{17} + x_{16}x_{17} + x_3x_{18} + x_4x_{18} + x_7x_{18} + x_8x_{18} + x_{10}x_{18} + x_{13}x_{18} + x_{15}x_{18} + x_{16}x_{18} + x_{17}x_{18} + x_{18} + x_1x_{19} + x_3x_{19} + x_5x_{19} + x_6x_{19} + x_7x_{19} + x_8x_{19} + x_9x_{19} + x_{11}x_{19} + x_{12}x_{19} + x_{17}x_{19} + x_{19} + x_3x_{20} + x_6x_{20} + x_7x_{20} + x_9x_{20} + x_{11}x_{20} + x_{16}x_{20} + x_{18}x_{20} + x_{19}x_{20},$$

$$y_{11} = P_{11}(x) = x_1x_2 + x_3x_2 + x_4x_2 + x_5x_2 + x_6x_2 + x_8x_2 + x_{11}x_2 + x_{12}x_2 + x_{16}x_2 + x_{17}x_2 + x_{19}x_2 + x_3x_4 + x_4 + x_1x_5 + x_3x_5 + x_4x_5 + x_5 + x_1x_6 + x_3x_6 + x_5x_6 + x_6 + x_4x_7 + x_6x_7 + x_7 + x_1x_8 + x_3x_8 + x_4x_8 + x_6x_8 + x_8 + x_1x_9 + x_4x_9 + x_5x_9 + x_7x_9 + x_9 + x_1x_{10} + x_3x_{10} + x_7x_{10} + x_8x_{10} + x_{10} + x_1x_{11} + x_3x_{11} + x_4x_{11} + x_5x_{11} + x_8x_{11} + x_1x_{12} + x_5x_{12} + x_6x_{12} + x_9x_{12} + x_{10}x_{12} + x_{11}x_{12} + x_1x_{13} + x_5x_{13} + x_6x_{13} + x_8x_{13} + x_9x_{13} + x_{10}x_{13} + x_{11}x_{13} + x_{12}x_{13} + x_{13} + x_4x_{14} + x_5x_{14} + x_7x_{14} + x_{11}x_{14} + x_{12}x_{14} + x_1x_{15} + x_5x_{15} + x_9x_{15} + x_{10}x_{15} + x_{11}x_{15} + x_{12}x_{15} + x_{13}x_{15} + x_3x_{16} + x_5x_{16} + x_6x_{16} + x_8x_{16} + x_{10}x_{16} + x_{11}x_{16} + x_{13}x_{16} + x_{16} + x_1x_{17} + x_3x_{17} + x_6x_{17} + x_7x_{17} + x_8x_{17} + x_9x_{17} + x_{10}x_{17} + x_{11}x_{17} + x_{13}x_{17} + x_{14}x_{17} + x_{16}x_{17} + x_{17} + x_3x_{18} + x_5x_{18} + x_9x_{18} + x_{10}x_{18} + x_{13}x_{18} + x_{14}x_{18} + x_{15}x_{18} + x_{18} + x_5x_{19} + x_7x_{19} + x_{11}x_{19} + x_{13}x_{19} + x_{16}x_{19} + x_{19} + x_3x_{20} + x_4x_{20} + x_5x_{20} + x_7x_{20} + x_9x_{20} + x_{11}x_{20} + x_{12}x_{20} + x_{13}x_{20} + x_{15}x_{20} + x_{16}x_{20} + x_{20},$$

$$y_{12} = P_{12}(x) = x_3x_2 + x_{11}x_2 + x_{13}x_2 + x_{14}x_2 + x_{16}x_2 + x_{19}x_2 + x_{20}x_2 + x_2 + x_1x_3 + x_3 + x_1x_4 + x_3x_4 + x_3x_5 + x_4x_5 + x_5 + x_1x_6 + x_1x_7 + x_4x_7 + x_6x_7 + x_7 + x_1x_8 + x_3x_8 + x_5x_8 + x_1x_9 + x_3x_9 + x_4x_9 + x_5x_9 + x_6x_9 + x_7x_9 + x_8x_9 + x_6x_{10} + x_8x_{10} + x_1x_{11} + x_3x_{11} + x_4x_{11} + x_7x_{11} + x_{10}x_{11} + x_{11} + x_1x_{12} + x_4x_{12} + x_5x_{12} + x_6x_{12} + x_7x_{12} + x_8x_{12} + x_9x_{12} + x_{10}x_{12} + x_{12} + x_1x_{13} + x_5x_{13} + x_{10}x_{13} + x_{12}x_{13} + x_3x_{14} + x_7x_{14} + x_9x_{14} + x_{12}x_{14} + x_{13}x_{14} + x_1x_{15} + x_3x_{15} + x_4x_{15} + x_6x_{15} + x_9x_{15} + x_{10}x_{15} + x_{11}x_{15} + x_{12}x_{15} + x_{14}x_{15} + x_1x_{16} + x_3x_{16} + x_5x_{16} + x_{10}x_{16} + x_{12}x_{16} + x_{14}x_{16} + x_1x_{17} + x_3x_{17} + x_5x_{17} + x_7x_{17} + x_8x_{17} + x_{11}x_{17} + x_{13}x_{17} + x_{14}x_{17} + x_{16}x_{17} + x_3x_{18} + x_6x_{18} + x_9x_{18} + x_{15}x_{18} + x_{17}x_{18} + x_1x_{19} + x_3x_{19} + x_4x_{19} + x_5x_{19} + x_6x_{19} + x_7x_{19} + x_9x_{19} + x_{10}x_{19} + x_{11}x_{19} + x_{13}x_{19} + x_{14}x_{19} + x_{18}x_{19} + x_1x_{20} + x_8x_{20} + x_{10}x_{20} + x_{11}x_{20} + x_{13}x_{20} + x_{14}x_{20} + x_{15}x_{20} + x_{17}x_{20} + x_{18}x_{20} + x_{20} + 1,$$

$$y_{13} = P_{13}(x) = x_2x_1 + x_4x_1 + x_5x_1 + x_6x_1 + x_7x_1 + x_9x_1 + x_{16}x_1 + x_{17}x_1 + x_{19}x_1 + x_{20}x_1 + x_1 + x_2x_3 + x_3 + x_3x_4 + x_3x_5 + x_4x_5 + x_3x_6 + x_4x_6 + x_5x_6 + x_3x_7 + x_6x_7 + x_4x_8 + x_8 + x_3x_9 + x_4x_9 + x_5x_9 + x_6x_9 + x_7x_9 + x_9 + x_2x_{10} + x_3x_{10} + x_5x_{10} + x_8x_{10} + x_9x_{10} + x_2x_{11} + x_3x_{11} + x_4x_{11} + x_5x_{11} + x_8x_{11} + x_9x_{11} + x_{10}x_{11} + x_{11} + x_2x_{12} + x_4x_{12} + x_5x_{12} + x_6x_{12} + x_8x_{12} + x_{10}x_{12} + x_2x_{13} + x_6x_{13} + x_7x_{13} + x_8x_{13} + x_{10}x_{13} + x_{11}x_{13} + x_3x_{14} + x_4x_{14} + x_6x_{14} + x_7x_{14} + x_8x_{14} + x_9x_{14} + x_{10}x_{14} + x_{11}x_{14} + x_{12}x_{14} + x_{13}x_{14} + x_{14} + x_3x_{15} + x_4x_{15} + x_7x_{15} + x_{10}x_{15} + x_{11}x_{15} + x_{13}x_{15} + x_{14}x_{15} + x_3x_{16} + x_4x_{16} + x_6x_{16} + x_{10}x_{16} + x_{12}x_{16} + x_{13}x_{16} + x_{14}x_{16} + x_{15}x_{16} + x_2x_{17} + x_3x_{17} + x_6x_{17} + x_8x_{17} + x_9x_{17} + x_{10}x_{17} + x_{12}x_{17} + x_6x_{18} + x_8x_{18} + x_9x_{18} + x_{10}x_{18} + x_{11}x_{18} + x_{12}x_{18} + x_{14}x_{18} + x_{15}x_{18} + x_{16}x_{18} + x_4x_{19} + x_9x_{19} + x_{10}x_{19} + x_{12}x_{19} + x_{13}x_{19} + x_{15}x_{19} + x_{19} + x_2x_{20} + x_3x_{20} + x_5x_{20} + x_6x_{20} + x_{11}x_{20} + x_{16}x_{20} + 1,$$

$$y_{14} = P_{14}(x) = x_1x_2 + x_5x_2 + x_8x_2 + x_{10}x_2 + x_{11}x_2 + x_{12}x_2 + x_{16}x_2 + x_{19}x_2 + x_{20}x_2 + x_2 + x_1x_3 + x_1x_4 + x_1x_5 + x_3x_5 + x_4x_5 + x_5 + x_1x_6 + x_5x_6 + x_6 + x_4x_7 + x_7 + x_1x_8 + x_4x_8 + x_7x_8 + x_8 + x_1x_9 + x_3x_9 + x_4x_9 + x_6x_9 + x_7x_9 + x_8x_9 + x_9 + x_1x_{10} + x_3x_{10} + x_4x_{10} + x_7x_{10} + x_8x_{10} + x_9x_{10} + x_1x_{11} + x_4x_{11} + x_7x_{11} + x_8x_{11} + x_9x_{11} + x_1x_{12} + x_5x_{12} + x_6x_{12} + x_8x_{12} + x_9x_{12} + x_{10}x_{12} + x_{11}x_{12} + x_{12} + x_1x_{13} + x_5x_{13} + x_6x_{13} + x_8x_{13} + x_{11}x_{13} + x_{12}x_{13} + x_1x_{14} + x_3x_{14} + x_4x_{14} + x_5x_{14} + x_6x_{14} + x_7x_{14} + x_{10}x_{14} + x_{11}x_{14} + x_{13}x_{14} + x_1x_{15} + x_3x_{15} + x_4x_{15} + x_5x_{15} + x_7x_{15} + x_8x_{15} + x_{11}x_{15} + x_{12}x_{15} + x_1x_{16} + x_3x_{16} + x_6x_{16} + x_8x_{16} + x_9x_{16} + x_{10}x_{16} + x_{11}x_{16} + x_{12}x_{16} + x_{14}x_{16} + x_{15}x_{16} + x_{16} + x_1x_{17} + x_3x_{17} + x_5x_{17} + x_7x_{17} + x_8x_{17} + x_{10}x_{17} + x_{11}x_{17} + x_{13}x_{17} + x_{15}x_{17} + x_1x_{18} + x_5x_{18} + x_6x_{18} + x_8x_{18} + x_{10}x_{18} + x_{14}x_{18} + x_{15}x_{18} + x_{16}x_{18} + x_{17}x_{18} + x_1x_{19} + x_3x_{19} + x_4x_{19} + x_5x_{19} + x_6x_{19} + x_8x_{19} + x_{10}x_{19} + x_{11}x_{19} + x_{12}x_{19} + x_{13}x_{19} + x_{14}x_{19} + x_{15}x_{19} + x_1x_{20} + x_3x_{20} + x_5x_{20} + x_6x_{20} + x_{12}x_{20} + x_{13}x_{20} + x_{15}x_{20} + x_{17}x_{20} + x_{18}x_{20} + x_{19}x_{20},$$

$$y_{15} = P_{15}(x) = x_4x_2 + x_8x_2 + x_9x_2 + x_{12}x_2 + x_{14}x_2 + x_{15}x_2 + x_{18}x_2 + x_2 + x_3 + x_3x_4 + x_4 + x_4x_5 + x_1x_6 + x_5x_6 + x_1x_7 + x_5x_7 + x_1x_8 + x_5x_8 + x_7x_8 + x_8 + x_1x_9 + x_3x_9 + x_4x_9 + x_5x_9 + x_7x_9 + x_8x_9 + x_9 + x_1x_{10} + x_5x_{10} + x_6x_{10} + x_9x_{10} + x_{10} + x_3x_{11} + x_7x_{11} + x_9x_{11} + x_{10}x_{11} + x_3x_{12} + x_4x_{12} + x_5x_{12} + x_6x_{12} + x_7x_{12} + x_8x_{12} + x_{11}x_{12} + x_5x_{13} + x_6x_{13} + x_{10}x_{13} + x_{12}x_{13} + x_{13} + x_8x_{14} + x_9x_{14} + x_{10}x_{14} + x_{11}x_{14} + x_{12}x_{14} + x_{13}x_{14} + x_3x_{15} + x_4x_{15} + x_5x_{15} + x_6x_{15} + x_8x_{15} + x_{10}x_{15} + x_1x_{16} + x_6x_{16} + x_9x_{16} + x_{10}x_{16} + x_{12}x_{16} + x_{14}x_{16} + x_{15}x_{16} + x_3x_{17} + x_4x_{17} + x_5x_{17} + x_7x_{17} + x_8x_{17} + x_{11}x_{17} + x_{12}x_{17} + x_{14}x_{17} + x_1x_{18} + x_7x_{18} + x_9x_{18} + x_{13}x_{18} + x_{17}x_{18} + x_3x_{19} + x_8x_{19} + x_9x_{19} + x_{11}x_{19} + x_{12}x_{19} + x_{15}x_{19} + x_{17}x_{19} + x_{19} + x_3x_{20} + x_4x_{20} + x_{11}x_{20} + x_{13}x_{20} + x_{16}x_{20} + x_{17}x_{20} + x_{18}x_{20} + x_{20},$$

$$y_{16} = P_{16}(x) = x_2x_1 + x_4x_1 + x_5x_1 + x_6x_1 + x_7x_1 + x_8x_1 + x_9x_1 + x_{10}x_1 + x_{12}x_1 + x_{13}x_1 + x_{14}x_1 + x_{16}x_1 + x_{18}x_1 + x_{20}x_1 + x_1 + x_2 + x_2x_3 + x_3 + x_2x_4 + x_3x_4 + x_4x_5 + x_5 + x_2x_6 + x_4x_6 + x_3x_7 + x_5x_7 + x_6x_7 + x_7 + x_5x_8 + x_7x_8 + x_2x_9 + x_2x_{10} + x_3x_{10} + x_4x_{10} + x_5x_{10} + x_6x_{10} + x_{10} + x_4x_{11} + x_7x_{11} + x_9x_{11} + x_{10}x_{11} + x_{11} + x_3x_{12} + x_5x_{12} + x_8x_{12} + x_9x_{12} + x_{12} + x_2x_{13} + x_3x_{13} + x_4x_{13} + x_6x_{13} + x_{11}x_{13} + x_4x_{14} + x_6x_{14} + x_9x_{14} + x_{10}x_{14} + x_{11}x_{14} + x_{12}x_{14} + x_{13}x_{14} + x_{14} + x_2x_{15} + x_3x_{15} + x_5x_{15} + x_6x_{15} + x_9x_{15} + x_{12}x_{15} + x_{13}x_{15} + x_{15} + x_3x_{16} + x_7x_{16} + x_8x_{16} + x_9x_{16} + x_{15}x_{16} + x_2x_{17} + x_4x_{17} + x_9x_{17} + x_{10}x_{17} + x_{11}x_{17} + x_{13}x_{17} + x_{14}x_{17} + x_2x_{18} + x_3x_{18} + x_7x_{18} + x_8x_{18} + x_{10}x_{18} + x_{14}x_{18} + x_{18} + x_3x_{19} + x_5x_{19} + x_7x_{19} + x_8x_{19} + x_9x_{19} + x_{10}x_{19} + x_{11}x_{19} + x_{12}x_{19} + x_{15}x_{19} + x_{16}x_{19} + x_{18}x_{19} + x_{19} + x_3x_{20} + x_8x_{20} + x_9x_{20} + x_{10}x_{20} + x_{11}x_{20} + x_{12}x_{20} + x_{13}x_{20} + x_{16}x_{20} + x_{18}x_{20} + 1,$$

$$y_{17} = P_{17}(x) = x_2x_1 + x_3x_1 + x_5x_1 + x_6x_1 + x_9x_1 + x_{10}x_1 + x_{11}x_1 + x_{12}x_1 + x_{13}x_1 + x_{14}x_1 + x_{15}x_1 + x_{18}x_1 + x_1 + x_2 + x_2x_4 + x_3x_4 + x_4 + x_2x_5 + x_3x_5 + x_4x_5 + x_3x_6 + x_5x_6 + x_4x_7 + x_3x_8 + x_5x_8 + x_6x_8 + x_7x_8 + x_2x_9 + x_3x_9 + x_4x_9 + x_5x_9 + x_7x_9 + x_8x_9 + x_9 + x_3x_{10} + x_4x_{10} + x_7x_{10} + x_8x_{10} + x_{10} + x_2x_{11} + x_3x_{11} + x_5x_{11} + x_6x_{11} + x_7x_{11} + x_9x_{11} + x_{11} + x_2x_{12} + x_6x_{12} + x_8x_{12} + x_9x_{12} + x_{11}x_{12} + x_{12} + x_3x_{13} + x_7x_{13} + x_8x_{13} + x_9x_{13} + x_{10}x_{13} + x_{11}x_{13} + x_{12}x_{13} + x_2x_{14} + x_3x_{14} + x_4x_{14} + x_5x_{14} + x_6x_{14} + x_8x_{14} + x_9x_{14} + x_{14} + x_4x_{15} + x_5x_{15} + x_7x_{15} + x_9x_{15} + x_{10}x_{15} + x_{11}x_{15} + x_{12}x_{15} + x_{13}x_{15} + x_{15} + x_3x_{16} + x_6x_{16} + x_8x_{16} + x_{10}x_{16} + x_{11}x_{16} + x_{15}x_{16} + x_3x_{17} + x_5x_{17} + x_6x_{17} + x_8x_{17} + x_9x_{17} + x_{12}x_{17} + x_{13}x_{17} + x_{15}x_{17} + x_{16}x_{17} + x_2x_{18} + x_6x_{18} + x_7x_{18} + x_{11}x_{18} + x_{12}x_{18} + x_{13}x_{18} + x_{16}x_{18} + x_{17}x_{18} + x_{18} + x_2x_{19} + x_4x_{19} + x_6x_{19} + x_7x_{19} + x_8x_{19} + x_9x_{19} + x_{13}x_{19} + x_{15}x_{19} + x_{16}x_{19} + x_{17}x_{19} + x_{18}x_{19} + x_4x_{20} + x_5x_{20} + x_8x_{20} + x_9x_{20} + x_{10}x_{20} + x_{12}x_{20} + x_{14}x_{20},$$

$$y_{18} = P_{18}(x) = x_1x_2 + x_3x_2 + x_4x_2 + x_7x_2 + x_{10}x_2 + x_{12}x_2 + x_{13}x_2 + x_{14}x_2 + x_{15}x_2 + x_{16}x_2 + x_{18}x_2 + x_{19}x_2 + x_1x_4 + x_4 + x_1x_5 + x_3x_5 + x_3x_6 + x_4x_6 + x_5x_6 + x_3x_8 + x_5x_8 + x_6x_8 + x_1x_9 + x_3x_9 + x_4x_9 + x_5x_9 + x_6x_9 + x_9 + x_1x_{10} + x_5x_{10} + x_7x_{10} + x_9x_{10} + x_{10} + x_3x_{11} + x_4x_{11} + x_5x_{11} + x_7x_{11} + x_{10}x_{11} + x_{11} + x_3x_{12} + x_4x_{12} + x_5x_{12} + x_7x_{12} + x_8x_{12} + x_9x_{12} + x_{11}x_{12} + x_8x_{13} + x_{10}x_{13} + x_{11}x_{13} + x_{13} + x_1x_{14} + x_3x_{14} + x_5x_{14} + x_6x_{14} + x_7x_{14} + x_8x_{14} + x_{11}x_{14} + x_{12}x_{14} + x_{14} + x_1x_{15} + x_3x_{15} + x_5x_{15} + x_7x_{15} + x_{12}x_{15} + x_{14}x_{15} + x_1x_{16} + x_4x_{16} + x_9x_{16} + x_{11}x_{16} + x_{14}x_{16} + x_{16} + x_1x_{17} + x_5x_{17} + x_6x_{17} + x_{10}x_{17} + x_{16}x_{17} + x_1x_{18} + x_3x_{18} + x_5x_{18} + x_6x_{18} + x_7x_{18} + x_8x_{18} + x_9x_{18} + x_{10}x_{18} + x_{13}x_{18} + x_{16}x_{18} + x_{18} + x_1x_{19} + x_3x_{19} + x_7x_{19} + x_8x_{19} + x_{10}x_{19} + x_{11}x_{19} + x_{12}x_{19} + x_{13}x_{19} + x_{14}x_{19} + x_{15}x_{19} + x_{18}x_{19} + x_{19} + x_4x_{20} + x_7x_{20} + x_9x_{20} + x_{10}x_{20} + x_{11}x_{20} + x_{13}x_{20} + x_{15}x_{20} + x_{16}x_{20} + x_{19}x_{20} + x_{20} + 1,$$

$$y_{19} = P_{19}(x) = x_3x_1 + x_4x_1 + x_5x_1 + x_6x_1 + x_7x_1 + x_9x_1 + x_{11}x_1 + x_{15}x_1 + x_{16}x_1 + x_{17}x_1 + x_{18}x_1 + x_{19}x_1 + x_{20}x_1 + x_1 + x_3 + x_3x_4 + x_4x_5 + x_5 + x_2x_6 + x_3x_6 + x_4x_6 + x_5x_6 + x_6 + x_6x_7 + x_7 + x_3x_8 + x_4x_8 + x_6x_8 + x_7x_8 + x_9 + x_2x_{10} + x_4x_{10} + x_5x_{10} + x_6x_{10} + x_{10} + x_2x_{11} + x_4x_{11} + x_6x_{11} + x_8x_{11} + x_9x_{11} + x_{10}x_{11} + x_2x_{12} + x_5x_{12} + x_7x_{12} + x_9x_{12} + x_{10}x_{12} + x_{12} + x_7x_{13} + x_9x_{13} + x_{12}x_{13} + x_3x_{14} + x_4x_{14} + x_6x_{14} + x_8x_{14} + x_{10}x_{14} + x_{12}x_{14} + x_{13}x_{14} + x_3x_{15} + x_6x_{15} + x_8x_{15} + x_9x_{15} + x_{12}x_{15} + x_{15} + x_2x_{16} + x_3x_{16} + x_4x_{16} + x_8x_{16} + x_9x_{16} + x_{10}x_{16} + x_{13}x_{16} + x_{15}x_{16} + x_2x_{17} + x_3x_{17} + x_4x_{17} + x_7x_{17} + x_{10}x_{17} + x_{11}x_{17} + x_{15}x_{17} + x_3x_{18} + x_4x_{18} + x_6x_{18} + x_7x_{18} + x_{12}x_{18} + x_{13}x_{18} + x_{15}x_{18} + x_{16}x_{18} + x_3x_{19} + x_4x_{19} + x_6x_{19} + x_9x_{19} + x_{11}x_{19} + x_{13}x_{19} + x_{16}x_{19} + x_{17}x_{19} + x_{18}x_{19} + x_5x_{20} + x_6x_{20} + x_7x_{20} + x_9x_{20} + x_{13}x_{20} + x_{14}x_{20} + x_{18}x_{20} + x_{19}x_{20} + x_{20},$$

$$y_{20} = P_{20}(x) = x_3x_1 + x_4x_1 + x_5x_1 + x_6x_1 + x_8x_1 + x_9x_1 + x_{13}x_1 + x_{15}x_1 + x_{20}x_1 + x_1 + x_3 + x_3x_4 + x_2x_5 + x_5 + x_3x_6 + x_4x_6 + x_5x_6 + x_2x_7 + x_3x_7 + x_6x_7 + x_7 + x_4x_8 + x_8 + x_2x_9 + x_3x_9 + x_9 + x_2x_{10} + x_3x_{10} + x_6x_{10} + x_7x_{10} + x_8x_{10} + x_{10} + x_3x_{11} + x_5x_{11} + x_9x_{11} + x_2x_{12} + x_4x_{12} + x_5x_{12} + x_6x_{12} + x_7x_{12} + x_9x_{12} + x_2x_{13} + x_3x_{13} + x_4x_{13} + x_7x_{13} + x_{10}x_{13} + x_{11}x_{13} + x_{12}x_{13} + x_2x_{14} + x_3x_{14} + x_4x_{14} + x_6x_{14} + x_{10}x_{14} + x_{11}x_{14} + x_{12}x_{14} + x_{13}x_{14} + x_5x_{15} + x_8x_{15} + x_9x_{15} + x_{12}x_{15} + x_{13}x_{15} + x_{15} + x_2x_{16} + x_8x_{16} + x_9x_{16} + x_{11}x_{16} + x_{12}x_{16} + x_{13}x_{16} + x_{14}x_{16} + x_{15}x_{16} + x_9x_{17} + x_{11}x_{17} + x_{14}x_{17} + x_{15}x_{17} + x_3x_{18} + x_7x_{18} + x_9x_{18} + x_{11}x_{18} + x_{16}x_{18} + x_{17}x_{18} + x_{18} + x_4x_{19} + x_5x_{19} + x_6x_{19} + x_7x_{19} + x_8x_{19} + x_9x_{19} + x_{11}x_{19} + x_{12}x_{19} + x_{13}x_{19} + x_{17}x_{19} + x_3x_{20} + x_6x_{20} + x_7x_{20} + x_8x_{20} + x_{10}x_{20} + x_{11}x_{20} + x_{12}x_{20} + x_{14}x_{20} + x_{16}x_{20} + x_{17}x_{20} + x_{19}x_{20} + x_{20} + 1.$$